\documentclass[11pt,letterpaper,twocolumn]{aastex63}
\usepackage{verbatim}
\usepackage{amsmath}

\usepackage{epsfig}
\usepackage{amsmath}
\usepackage{natbib}
\usepackage{mathtools}

\begin{document}
 
\title{A Scatter of Light from a Polarized World}
\author{Sloane J. Wiktorowicz}
\affiliation{Remote Sensing Department, The Aerospace Corporation, 2310 E. El Segundo Blvd., El Segundo, CA 90245}
\email{s.wiktorowicz@aero.org}

\author{Pushkar Kopparla}
\affiliation{Solafune, Inc., Solix Shibuya 401, 3-chōme-6-15, Shibuya, Shibuya City, Tokyo, 150-0002}

\author{Jiazheng Li}
\affiliation{Space Science Institute, Macau University of Science and Technology, Avenida Wai Long, Taipa, Macau, China}

\author{Yuk L. Yung}
\affiliation{Division of Geological and Planetary Sciences, California Institute of Technology, 1200 E. California Blvd., Pasadena, CA 91125}

\date{\today}

\begin{abstract}

Many known exoplanets harbor clouds, which lead to degeneracies in spectroscopic models between particle composition and size. Polarimetry, however, provides independent assessment. Here we report the $7.2\sigma$ discovery of linearly polarized, scattered light from the hot Jupiter HD 189733b in $B$ band (390 to 475 nm) peaking near quarter phase with $\Delta p=40.9 \pm 7.1$ ppm. Polarization measurements, obtained with the POLISH2 polarimeter at both Gemini North and the Lick Observatory 3-m, are best explained by silicate (SiO$_2$ or MgSiO$_3$) particles with effective radius $r_\text{eff}=0.038^{+0.047}_{-0.023}$ $\mu$m ($90\%$ confidence). This is broadly consistent with results from both Hubble transmission spectroscopy and JWST secondary eclipse spectroscopy suggesting small, SiO$_2$ scattering particles. It is difficult to reconcile large polarization and moderate Hubble secondary eclipse depth via pure Rayleigh, silicate, or MnS scatterers. The measured polarization of HD 189733b is detected with such high confidence that we place a $2\sigma$ lower limit on its $B$ band geometric albedo of $A_g>0.26$ with a preferred value of $A_g=0.6$. This is larger than the prior estimate of $A_g=0.226\pm0.091$ from Hubble secondary eclipse photometry, and it presents HD 189733b as one of the most reflective known exoplanets in $B$ band. It also validates Rayleigh scattering from the exoplanet, as opposed to starspot contamination, as the cause of HD 189733's blue optical slope in transmission spectroscopy. Assuming other known exoplanets harbor atmospheres like HD 189733b, we model dozens to be detectable with at least $5\sigma$ confidence after one week of Gemini time each. \\

\end{abstract}


\section{Introduction}\label{sec1}

Our understanding of the chemical constituents of exoplanet atmospheres comes mainly from transmission spectroscopy. Such measurements tend to be obtained by Hubble during transit, when a thin ring of transmitted starlight from the host is observed to encircle the night side of the planet. On the other side of the orbit, secondary eclipse allows direct scattered or emitted light from the dayside of the exoplanet to be detected as it is occulted by the host star and subsequently reappears. Full-orbit, scattered light measurements were enabled by Kepler's groundbreaking coverage, and ground-based spectroscopic observations of exoplanet daysides have detected molecules in their atmospheres. This technique may be performed at almost arbitrary location in the exoplanet's orbit, and it is possible even for the essentially ignored population of non-transiting exoplanets.

Ironically, while bypassing terrestrial clouds that plague ground-based observatories, Hubble has run headlong into the pervasive presence of clouds in exoplanets, which obscures signatures of underlying chemical species \citep{Iyer2016, Sing2016}. Dramatic in its ability to dissect clouds and hazes in the solar system \citep{Lyot1929, Coffeen1969, Hansen1974, Tomasko1982, West1983, Smith1984, Tomasko1984}, polarimetric detection of exoplanets has not yet been conclusive \citep{Berdyugina2008, Lucas2009, Wiktorowicz2009, Berdyugina2011, Wiktorowicz2015_189, Bott2016, Bott2018, Bailey2020, Bailey2021}. This is due to the difficulty in maintaining proper calibration. Indeed, while ground-based transit and dayside observations typically conclude during a single night \citep{Birkby2013}, rigorous calibration of exoplanet polarimetry must encompass repeated measurements during times of high and low signature throughout the planetary orbit. This necessitates careful calibration at the part-per-million level maintained over days, weeks, and potentially months and years.

\section{Methods}
\subsection{Models}

We utilize our multiple scattering, photopolarimetric model \citep{Kopparla2016, Kopparla2018}, based on the vector radiative transfer model VLIDORT \citep{Spurr2006}, to predict scattered light photopolarimetry of HD 189733b at all phases of its orbit. We model peak polarization to be $\Delta p \sim 26$ ppm for pure Rayleigh scatterers due to contamination from direct light of the host star, which is consistent with the literature \citep{Bailey2018}. This levies stringent night-to-night calibration requirements on ground-based instruments.

Hubble transmission spectroscopy of HD 189733b, which probes significantly shallower in the atmosphere than does full-phase scattered light polarimetry, suggests the presence of Rayleigh scattering clouds composed of small particles \citep{Lecavelier2008, Pont2008}. Additionally, the wavelength dependent slope of flux transmitted through HD 189733b is significantly steeper than expected for pure Rayleigh scattering (so-called ``super-Rayleigh slope"), which is potentially caused by strong mixing in the exoplanet atmosphere or by starspot contamination \citep{McCullough2014, Welbanks2019, Gardner2023}.

Model pressure-temperature profiles suggest SiO$_2$, MgSiO$_3$, and MnS to be thermodynamically stable in the upper atmosphere of HD 189733b \citep{Pinhas2017}, so we generate polarization models containing pure Rayleigh, SiO$_2$, MgSiO$_3$, and MnS particles. The pure Rayleigh scattering model utilizes a gaseous scattering layer with optical depth $\tau = 10^4$ overlying a perfectly reflective, Lambertian surface. We vary the single scattering albedo, which is directly related to the geometric albedo $A_g$, to vary the linear polarization phase curve to compare to observations. Clouds are assumed to be composed entirely of SiO$_2$, MgSiO$_3$, or MnS particles, whose refractive indices in $B$ band ($\lambda_c = 440$ nm) are $1.48 + 0.0001i$, $1.58 + 0.002i$ and $3.14 + 0.05 i$, respectively \citep{Kitzmann2018}. While other species may be stable at the temperature-pressure conditions probed by optical observations, we limit our models to the leading candidates above.

Cloud particles are modeled to be spherical with a log-normal size distribution and variance of 0.1 times the effective particle radius $r_\text{eff}$. We expect this to be a reasonable approximation even for solid particles. For SiO$_2$, MgSiO$_3$, and MnS models, effective particle radii are varied from $r_\text{eff} = 0.01$ to 0.1 $\mu$m in 0.01 $\mu$m increments and from 0.2 to 1.0 $\mu$m in 0.1 $\mu$m increments. This enables a direct test of prior results suggesting that Rayleigh scattering dominates in HD 189733b at these wavelengths, as the Rayleigh-like scattering of small particles has fundamentally different photopolarimetric signatures from the Mie scattering of larger particles.

Exoplanet polarization models utilize stellar and planetary parameters listed in Table \ref{stardata}. The quantity $(R_p / a)^2$ determines the fraction of outgoing stellar flux that is intercepted by the planet's disk, and both photometric and polarimetric phase curves are scaled by this factor. Updated values from the community, in particular the effective planetary radius to scattered light in $B$ band, have a large effect on polarization models.

\begin{deluxetable*}{cccccccccc}
\tabletypesize{\normalsize}
\tablecaption{HD 189733 System Parameters}
\tablewidth{0pt}
\tablehead{
\colhead{Stellar Parameter} & \colhead{Value} & \colhead{Reference}}
\startdata
Right Ascension				& 20 00 43.7129433648	& \citet{Gaia2020} \\
Declination					& +22 42 39.073143456	& \citet{Gaia2020} \\
$B$ magnitude					& 8.578				& \citet{Koen2010} \\
Distance						& 19.8 pc				& \citet{Gaia2020} \\
Constant Stokes $q_*$			& $36.4 \pm 3.6$ ppm	& This work \\
Constant Stokes $u_*$			& $44.0 \pm 3.3$ ppm	& This work \\
Variable Stokes $\Delta v_*$		& $67 \pm 19$ ppm		& This work \\
Constant Stokes $p_*$			& $57.0 \pm 3.4$ ppm	& This work \\
Constant orientation $\Theta_*$	& $25.2^\circ \pm 1.8^\circ$	& This work \\
\hline
Planetary Orbital Parameter 		& Value \\
\hline
Eccentricity $e$				& 0	& \citet{Bouchy2005} \\
Inclination $i$					& $85.710^\circ \pm 0.024^\circ$	& \citet{Agol2010} \\
Argument of periastron $\omega$	& $90^\circ$	& \citet{Bouchy2005}  \\
Time of periastron $T_\text{peri}$	& MJD 54278.936714	& \cite{Agol2010} \\
Planetary radius $R_p$			& 1.138 $R_J$	& \citet{Torres2008} \\
Semimajor axis $a$				& 0.03100 au	& \citet{Butler2006} \\
Orbital period					& 2.21857567 d	& \citet{Agol2010} \\
Longitude of ascending node $\Omega$	& $13^\circ \pm 12^\circ$ ($90\%$ conf.)	& This work
\label{stardata}
\enddata
\end{deluxetable*}

\subsection{Observations}
\subsubsection{Overview}

The POLISH2 aperture-integrated polarimeter obtained 50 nights of $B$ band observations (390 to 475 nm, $\lambda_c = 419$ nm) of the transiting hot Jupiter host star HD 189733 at the Lick Observatory Shane 3-m telescope between UT 21 Jul 2011 and 14 Jul 2014 \citep{Wiktorowicz2015_189}. Here, we recalibrate these observations based on updated methodology \citep{Wiktorowicz2023}. Additionally, we present HD 189733 observations obtained with POLISH2 at the straight Cassegrain focus of the Gemini North 8-m telescope on UT 2 to 7 Aug 2018. Observations on only 3, 5, and 7 Aug were usable due to weather, and the dataset is dominated by observations on 7 Aug. While observations at Gemini North are limited in orbital phase, the Lick 3-m were obtained throughout the 2.22 d orbital period of the planet \citep{Agol2010}. This is because unlike transit and secondary eclipse signatures, polarization signatures may appear at nearly any orbital phase based on scattering particle size and index of refraction \citep{Seager2000}.

Fractional linear polarization $p$ is equal to the quadrature sum of polarization observables, which are given by Stokes parameters $q = Q/I$ and $u = U/I$. Stokes $Q$ represents the flux difference between photons with electric fields oscillating in the North-South minus East-West directions, and Stokes $U$ represents the flux difference for the Northeast-Southwest minus Northwest-Southeast directions. Fractional polarization $q$ and $u$ are calculated by normalizing Stokes $Q$ and $U$ by total photon flux $I$, which negates minor atmospheric transparency variations. In addition to linear polarization $q$, $u$, and $p$, the POLISH2 polarimeter simultaneously measures circular polarization (Stokes $v = V/I$) by the use of two photoelastic modulators instead of rotating waveplates. However, our models and the literature \citep{Rossi2018} suggest that disk-integrated circular polarization of scattered light from homogeneous exoplanet atmospheres will be zero.

Our models are consistent with others in the literature predicting Rayleigh-scattered polarization to be oriented perpendicular to the star-exoplanet-observer ``scattering plane," where $q^\prime \neq 0$ and $u^\prime = 0$ in this reference frame \citep{Seager2000, Bailey2018}. The planetary orbital angular momentum vector is aligned with $+q^\prime$, which is rotated in the plane of the sky from celestial $+q$ by the angle $\Omega + 270^\circ$ \citep{Wiktorowicz2009}. Here, $\Omega$ is the longitude of the ascending node of the orbit, and we rotate $(q^\prime, u^\prime)$ to the celestial reference frame $(q,u)$ by the following:

\begin{subequations}
\begin{align}
q & = - q^\prime \cos{2 \pi \Omega} + u^\prime \sin{2 \pi \Omega} \\
u & = - q^\prime \sin{2 \pi \Omega} - u^\prime \cos{2 \pi \Omega}.
\end{align}
\end{subequations}

For each of the atmospheric constituents (pure Rayleigh, SiO$_2$, MgSiO$_3$, and MnS particles), we perform a $\chi^2$ grid search against measured Stokes $q$ and $u$ by varying single scattering albedo (for pure Rayleigh scattering) or particle radius (for SiO$_2$, MgSiO$_3$, and MnS particles) while also varying $\Omega$ from $0^\circ$ to $179^\circ$ in $1^\circ$ increments. This is because the orientation of linear polarization rewraps after $180^\circ$ of rotation.

\begin{figure*}
\centering
\includegraphics[width=0.75\textwidth]{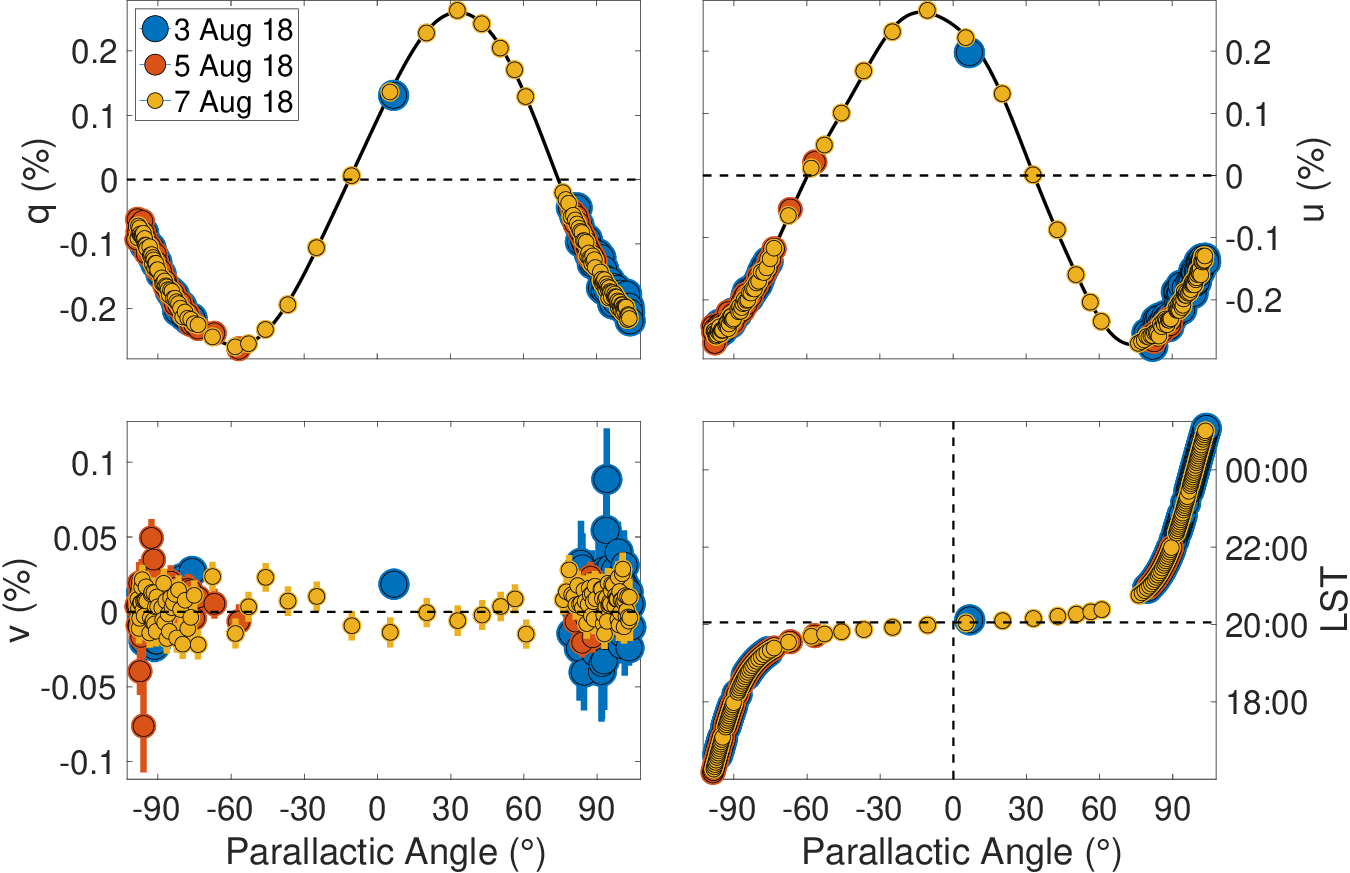}
\caption{Raw $B$ band polarization measurements of HD 189733 obtained at Gemini North on UT 1 to 7 Aug 2018, which are dominated by observations on the last night (gold points). \textit{Top left}: Stokes $q$ (linear polarization) modulation as a function of parallactic angle. Telescope linear polarization is modulated by parallactic angle with amplitude $\Delta (q_\text{TP},u_\text{TP}) = 0.2620 \pm 0.0034\%$ \citep{Wiktorowicz2023}, where the best fit curve from Equation \ref{eq:gemq} is shown in black. Uncertainties in most data points are smaller than the size of the data points. \textit{Top right}: Same as at top left but for Stokes $u$ (linear polarization, Equation \ref{eq:gemu}). \textit{Bottom left}: Same as at top left but for Stokes $v$ (circular polarization, Equation \ref{eq:gemv}). \textit{Bottom right}: Temporal variation of parallactic angle, which rotates rapidly as the target crosses the meridian (black dashed lines) around 20:00 LST. This rapid meridian crossing provides the bulk of the telescope polarization modulation necessary for accurate calibration.}
\label{gemrawdat}
\end{figure*}

\citet{Wiktorowicz2023} document the POLISH2 observation sequence and data reduction procedures in detail. Briefly, target stars are acquired by the telescope guide camera and sent to the POLISH2 aperture. Fine centering on POLISH2's 5 mm field stops (FOV of 8 arcsec diameter at Gemini North and 19 arcsec at the Lick 3-m) is accomplished by the POLISH2 on-axis guide camera. Polarization data are then recorded in target-sky-target ``triplets" with 30 sec duration per nod. Such frequent sky subtraction enables high accuracy sky polarization subtraction even when the background sky is strongly polarized by moonlight. Synchronous demodulation of the photomultiplier detector signals with respect to active polarization optics (dual photoelastic modulators) enables part-per-million polarization sensitivity to be obtained.

\subsubsection{Gemini North Run}

In a break with the tradition followed by essentially all part-per-million polarimetry to date, where telescope polarization is measured from a set of weakly polarized stars \citep{Hough2006, Wiktorowicz2008, Lucas2009, Wiktorowicz2009, Berdyugina2011, Bailey2015, WiktorowiczNofi2015, Wiktorowicz2015_189, Bott2016, Bailey2017, Bailey2020, Cotton2020, Marshall2020, Bailey2021}, we exploit the advantages offered by self-calibration of telescope polarization on alt-az telescopes using the science target \citep{Wiktorowicz2014, Millar-Blanchaer2020, Wiktorowicz2023}. This is because non-Gaussian systematic effects dominate at the part-per-million level. Long-predicted effects, such as in-band shifts in wavelength-dependent telescope polarization due to target star spectral type \citep{Hough2006}, overwhelm the exoplanet signal \citep{Bailey2020}. Indeed, it has been shown at Gemini North that even though any star enables highly accurate measurement of telescope polarization, the in-band estimate from star to star may vary by at least $0.04\% = 400$ ppm \citep{Wiktorowicz2023}. Thus, even the choice of telescope polarization calibrator necessarily biases in-band telescope polarization calibration of the science target.

\begin{figure*}
\centering
\includegraphics[width=0.75\textwidth]{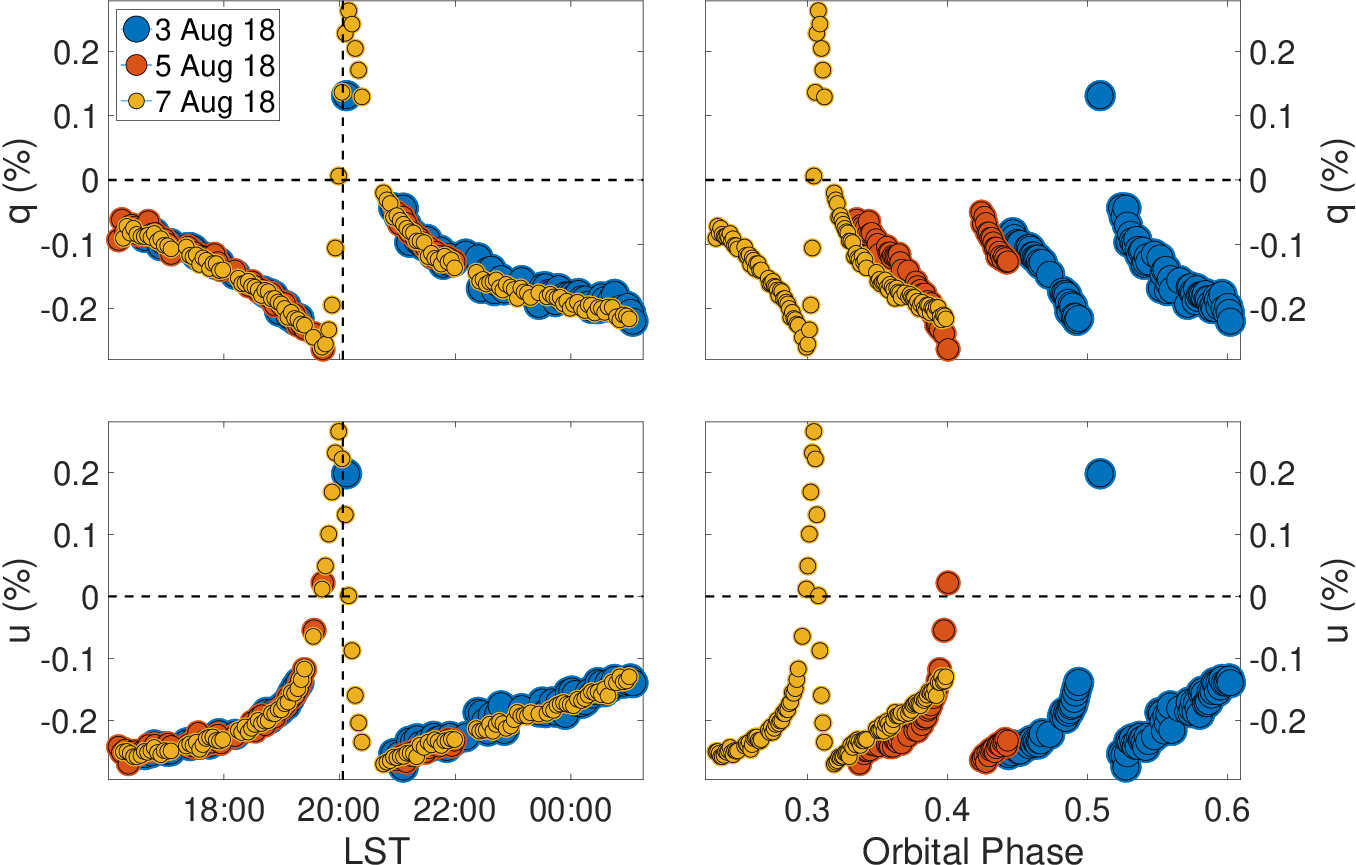}
\caption{\textit{Top left}: Stokes $q$ data from Figure \ref{gemrawdat} plotted as a function of LST, whose heartbeat nature is due to telescope polarization. Polarization modulates rapidly as the target crosses the meridian (vertical dashed line). \textit{Top right}: Stokes $q$ data as a function of HD 189733b orbital phase, where orbital phase 0 and 1 represent midtransit (planet between host star and observer) and orbital phase 0.5 represents the center of the secondary eclipse (host star between planet and observer). \textit{Bottom}: Same as at top but for Stokes $u$.}
\label{gempolLST}
\end{figure*}

Raw POLISH2 observations of HD 189733b obtained in $B$ band at Gemini North on UT 3, 5, and 7 Aug 2018 are shown in Figure \ref{gemrawdat}. Each data point represents 205 sec of data, and median uncertainties are $\sigma_q = 45$ and $\sigma_u = 47$ ppm. Since Gemini North is an alt-az telescope, a choice must be made to power the Cassegrain de-rotator on (so-called Follow mode) or off (Fixed mode). With the de-rotator powered on, the telescope primary and secondary mirrors rotate with parallactic angle $\phi$, and the instrument perceives Celestial North to be stationary. However, Figure \ref{gemrawdat} shows that nonzero linear polarization from the telescope, of order 100 ppm at most telescopes and dominating any exoplanet signature by at least one order of magnitude, rotates with parallactic angle according to the following \citep{Wiktorowicz2023}:

\begin{subequations}
\begin{align}
\label{eq:gemq}
q(\phi,t) &= q_* + q_\text{planet}(t) + a_\text{TP} \cos \omega \phi + b_\text{TP} \sin \omega \phi \\
\nonumber & + a_\text{TP2} \cos 2\omega \phi + b_\text{TP2} \sin 2\omega \phi \\
\label{eq:gemu}
u(\phi,t) &= u_* + u_\text{planet}(t) + c_\text{TP} \cos \omega \phi + d_\text{TP} \sin \omega \phi \\
\nonumber & + c_\text{TP2} \cos 2\omega \phi + d_\text{TP2} \sin 2\omega \phi \\
\label{eq:gemv}
v(t) & = v_*(t) + v_\text{TP} \\
\label{eq:tpq}
q_\text{TP} & = \sqrt{a_\text{TP}^2 + b_\text{TP}^2} \\
\label{eq:tpu}
& = \sqrt{c_\text{TP}^2 + d_\text{TP}^2} = u_\text{TP}.
\end{align}
\end{subequations}

\noindent Here, $\phi$ represents parallactic angle; stellar polarization is given by $q_*$, $u_*$ (linear polarization), and $v_*(t)$ (time-variable circular polarization, Section \ref{circsec}); time-variable exoplanet polarization is given by $q_\text{planet}(t)$ and $u_\text{planet}(t)$; exoplanet circular polarization $v_\text{planet}(t) = 0$ according to our models; and variables with subscript TP are various linear and circular polarization components of telescope polarization. Since circular polarization is invariant under rotation, separation of stellar from telescope circular polarization requires the same calibration procedure as for linear polarization calibration at the equatorial Lick 3-m telescope (Section \ref{licksec}).

With the Cassegrain de-rotator powered off, the reverse situation occurs: telescope polarization is fixed in the instrument frame, but the polarization vector of the target rotates with parallactic angle in the instrument frame. Given that the exoplanet signature is also time variable, this method would require fitting measured polarization to sinusoids in parallactic angle whose coefficients themselves vary in time based on the unknown polarization phase curve of the exoplanet. It is therefore much less complicated to power the de-rotator on and subtract sinusoidal telescope polarization to uncover intrinsic modulation due to the exoplanet.

\begin{deluxetable*}{cccccccccc}
\tabletypesize{\normalsize}
\tablecaption{Calibrated HD 189733b Measurements}
\tablewidth{0pt}
\tablehead{
\colhead{Telescope} & \colhead{Orbital Phase} & \colhead{$q$ (ppm)} & \colhead{$u$ (ppm)}	& \colhead{$v$ (ppm)}	& \colhead{$p$ (ppm)}	& \colhead{$\Theta$ ($^\circ$)} }
\startdata
Lick 3-m	& 0.052(50)	& $-$8(11)     	 & $-$4.9(9.7)  	 & 14(21)    	 & 7.6(7.2)  	 & 106(46)    \\
$\cdots$	& 0.155(50)	& 0.5(9.4)     	 & 12.9(8.5)    	 & 58(19)    	 & 12.9(8.5) 	 & 44(25)     \\
$\cdots$	& 0.239(50)	& $-$22(12)    	 & $-$0(10)     	 & 0(23)     	 & 22(12)    	 & 90(15)     \\
$\cdots$	& 0.352(50)	& $-$30(13)    	 & $-$31(11)    	 & 24(25)    	 & 42(12)    	 & 113.3(8.3) \\
$\cdots$	& 0.454(50)	& $-$1(11)     	 & $-$9(10)     	 & 53(22)    	 & 8.7(8.7)  	 & 133(57)    \\
$\cdots$	& 0.540(50)	& 4(10)        	 & 2.8(9.1)     	 & 66(20)    	 & 4.3(4.0)  	 & 16(90)     \\
$\cdots$	& 0.646(50)	& 4(13)        	 & 37(12)       	 & 38(26)    	 & 37(12)    	 & 42(10)     \\
$\cdots$	& 0.744(50)	& $-$9(11)     	 & $-$3.8(9.8)  	 & $-$15(21) 	 & 8.4(8.3)  	 & 102(45)    \\
$\cdots$	& 0.851(50)	& $-$23(11)    	 & $-$7.4(9.8)  	 & 23(22)    	 & 23(11)    	 & 99(13)     \\
$\cdots$	& 0.953(50)	& 8.0(9.9)     	 & $-$4.2(8.9)  	 & 67(19)    	 & 7.6(7.4)  	 & 166(42)    \\
\hline
Gemini 8-m	& 0.262(28)	& $-$27.8(6.7) 	 & $-$0.2(7.0)  	 & $-$8(15)  	 & 27.8(6.7) 	 & 90.2(7.5)  \\
$\cdots$	& 0.318(27)	& $-$37.6(7.0) 	 & $-$22.7(7.3) 	 & 18(16)    	 & 43.3(7.1) 	 & 105.6(4.8) \\
$\cdots$	& 0.374(25)	& $-$28.3(7.3) 	 & $-$22.2(7.6) 	 & 37(16)    	 & 35.2(7.4) 	 & 109.0(6.1) \\
$\cdots$	& 0.468(26)	& $-$1.9(5.9)  	 & $-$0.8(5.6)  	 & $-$  	 	& 1.8(1.8)  	 & 101(90)         
\label{lickgemdat}
\enddata
\end{deluxetable*}

Fitting Gemini North POLISH2 polarization measurements in Figure \ref{gemrawdat} to Equations \ref{eq:gemq} and \ref{eq:gemu} shows that they are dominated by gargantuan linear telescope polarization with $p_\text{TP} = 0.2620 \pm 0.0034\% = 2620 \pm 34$ ppm, which is an order of magnitude larger than at most telescopes. At the Lick 3-m for example, $p_\text{TP} \sim 0.01\% = 100$ ppm in $B$ band \citep{Wiktorowicz2023}. In spite of this, we use Equations \ref{eq:gemq} to \ref{eq:tpu} to subtract telescope polarization and uncover intrinsic modulation from the exoplanet with high accuracy (Section \ref{sec2a}). Stokes $q$ and $u$ modulation due to HD 189733b lie at the $\Delta (q,u) \sim 35$ ppm level and are a factor of $\sim 75$ times weaker than the contribution from telescope polarization. The combination of stellar and telescope circular polarization has a time average of $v = 27.7 \pm 5.4$ ppm and will be discussed in Section \ref{circsec}.

Telescope polarization measurements are repeatable from night to night, because the same parallactic angle locus is traced as a function of LST (Figure \ref{gemrawdat}, bottom right). The speed with which raw polarization measurements are modulated by telescope polarization is illustrated in Figure \ref{gempolLST}. Stokes $q$ and $u$ rapidly rotate as HD 189733 crosses the meridian, where parallactic angle $\phi = 0^\circ$. Indeed, the bottom right panel of Figure \ref{gemrawdat} shows that most of the sinusoidal modulation from telescope polarization occurs during the one-hour period as the target crosses the meridian, where the alt-az telescope mirrors rotate about the optical axis with a peak angular velocity of $0.08^\circ/s$. This provides relatively dense sampling of telescope linear polarization in parallactic angle, which is crucial for high accuracy fits to telescope polarization that enable exoplanet modulation to be uncovered.

\subsubsection{Lick 3-m Campaign}
\label{licksec}


Telescope polarization calibration at an equatorial telescope, like the Lick 3-m, is significantly more complicated than at an alt-az telescope. This is because the telescope, stellar, and exoplanet contributions are not easily separated like they are at alt-az telescopes. For the Lick campaign, a large set of weakly polarized stars is observed over multiple, week-long runs, and run-to-run variations in measured polarization are correlated from star to star with the assumption that variations are due to the telescope. Mean polarization of the stellar calibrator sample is assumed to be zero. This enables very high polarization sensitivity, the ability to measure change, as Lick 3-m telescope polarization is measured to be variable at the 10 ppm level or less over a ten-year period \citep{Wiktorowicz2023}.


During secondary eclipse at orbital phase 0.5, the exoplanet is completely occulted by the star. Since the exoplanet's contribution to system polarization must be zero here, we subtract measured Stokes $q$ and $u$ from both telescopes by their measurements near secondary eclipse. This static offset of $(q_*, u_*) \sim 40$ ppm is the time-averaged stellar linear polarization (Table \ref{stardata}). Table \ref{lickgemdat} lists intrinsic HD 189733 system polarization, measured at both the Lick 3-m and Gemini North, after subtraction of telescope and stellar polarization. Midtransit occurs at orbital phases 0 and 1, and the secondary eclipse center occurs at orbital phase 0.5. Quantities in parenthesis indicate the $1 \sigma$ uncertainty in the last two digits of the parameter. For orbital phase entries, quantities in parenthesis indicate half of the bin duration.

\begin{figure}
\centering
\includegraphics[width=0.47\textwidth]{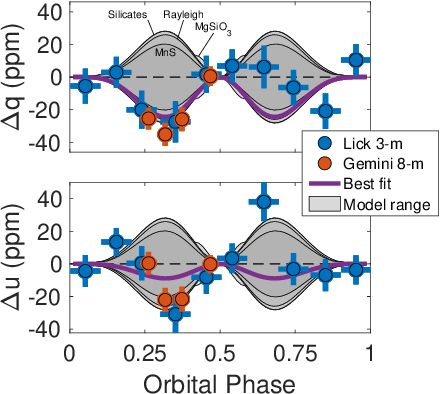}
\caption{\textit{Top}: POLISH2 $B$ band Stokes $q$ measurements obtained at the Lick 3-m (blue points) and Gemini North (red points) compared to the best fit SiO$_2$ scattering model ($r_\text{eff} = 0.04$ $\mu$m, $\Omega = 13^\circ$, solid purple curve). The full suite of pure Rayleigh, SiO$_2$, MgSiO$_3$ (labeled ``Silicates"), and MnS scattering models are also shown (grey regions). No models produce a significant rainbow feature that may explain outliers at orbital phases 0.65 or 0.85. \textit{Bottom}: Same as at top but for Stokes $u$. Measured peak linear polarization $\Delta p$ is $53.4 \pm 3.8\%$ larger than expected from the most polarized models.}
\label{allmodels}
\end{figure}

\section{Results}\label{sec2}
\subsection{Detection}\label{sec2a}

Calibrated observations from Table \ref{lickgemdat} are presented in Figure \ref{allmodels} along with the full range of pure Rayleigh, SiO$_2$, MgSiO$_3$, and MnS models. We discover repeatable signatures in both Stokes $q$ and $u$ that are not only phase locked to the orbital period of the exoplanet, but also peak and vanish at orbital phases consistent with scattering by small particles. Unlike prior reports of scattered light detection from this exoplanet with peak linear polarization of $100 < \Delta p < 200$ ppm \citep{Berdyugina2008, Berdyugina2011}, which is four to seven times larger than the most polarized, physically plausible multiple scattering models, the $\Delta p = 40.9 \pm 7.1$ ppm modulation detected in this work is $53.4 \pm 3.8\%$ larger than such models. Indeed, the large peak linear polarization from prior reports would have been detectable with POLISH2 \citep{Wiktorowicz2015_189}, HIPPI \citep{Bott2016}, and HIPPI-2 \citep{Bailey2021}. While peak degree of polarization $\Delta p$ in prior reports is invalidated by these subsequent investigations, the longitude of the ascending node measured in this work ($\Omega = 13^\circ \pm 12^\circ$) is well within the uncertainties in the original \citep[$\Omega = 16^\circ \pm 8^\circ$]{Berdyugina2008} and updated reports \citep[$\Omega = 14^\circ \pm 6^\circ$]{Berdyugina2011}.

Notably, peak $\Delta p = 40.9 \pm 7.1$ ppm in this work lies within the 99\% confidence upper limit of $\Delta p < 79$ ppm from POLISH \citep{Wiktorowicz2009}, the 99.7\% confidence upper limit of $\Delta p < 60$ ppm from POLISH2 \citep{Wiktorowicz2015_189}, and the best-fit peak of $\Delta p = 29 \pm 16$ ppm from HIPPI \citep{Bott2016}. While \citet{Bailey2021} suggest that HIPPI-2 data cannot rule out $\Delta p \sim 20$ ppm modulation, it is not obvious that their Stokes $q$ and $u$ data are capable of ruling out the larger, $| \Delta q | =  35.2 \pm 7.0$ ppm and $| \Delta u | =  31 \pm 11$ ppm modulation observed in this work.

However, the previously reported $B$ band geometric albedo upper limit of $A_g < 0.40$ from POLISH2 \citep[99.7\% confidence]{Wiktorowicz2015_189} is clearly invalidated by the observations in this work. As stated in Section \ref{sec1}, calibration is key to achieving the Kepler/CHEOPS-like precision and accuracy necessary to detect polarized, scattered light from close-in exoplanets from the ground. The detection in this work is due both to extensive POLISH2 recalibration \citep{Wiktorowicz2023}, which enabled Lick 3-m POLISH2 data \citep{Wiktorowicz2015_189} to uncover HD 189733b modulation (Figure \ref{allmodels}), and to the powerful polarization self-calibration afforded by alt-az telescopes such as Gemini North. While the amount of Gemini North data is limited, it is clear that the sensitivity and accuracy of POLISH2 measurements at Gemini North are far superior to those obtained at the Lick 3-m (Figure \ref{allmodels}), in spite of the fact that telescope polarization is both an order of magnitude larger at Gemini and strongly wavelength dependent \citep{Bailey2020}. Indeed, it is the Gemini North campaign that validates the Lick 3-m modulation and not the other way around. For example, a $3.9 \sigma$ outlier in Lick 3-m Stokes $u$ observations obtained near orbital phase 0.65 appears to suffer from a $\Delta u = 46$ ppm systematic effect (Figure \ref{allmodels}).

\begin{figure}
\centering
\includegraphics[width=0.47\textwidth]{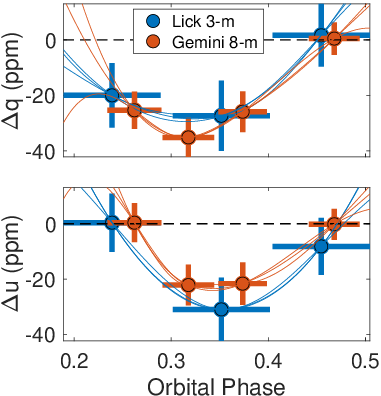}
\caption{Same as in Figure \ref{allmodels} but zoomed in to the orbital phase range where Gemini North and Lick 3-m data overlap. Gemini and Lick data reject a zero-polarization model with $7.2 \sigma$ and $2.9 \sigma$ confidence, respectively. Spline, Matlab \texttt{pchip}, and Matlab \texttt{makima} interpolations of the Lick 3-m and Gemini North data are given by blue and red curves, respectively.}
\label{allmodels_zoom}
\end{figure}

A constant, $\Delta q = \Delta u = 0$ null result is rejected with $7.2 \sigma$ confidence over the orbital phases where Gemini North and Lick 3-m data overlap (Figure \ref{allmodels_zoom} and Table \ref{conf}). The Gemini North data alone reject a null result with $7.2 \sigma$ confidence. We focus on this orbital phase overlap region because an unknown amount of Lick 3-m data obtained outside the Gemini North coverage appears to be subject to non-Gaussian systematic effects (i.e., the $3.9 \sigma$ outlier in Stokes $u$ at orbital phase 0.65 in Figure \ref{allmodels}). The standard approach of using Gaussian Process algorithms to account for such outliers is clearly not valid for data harboring non-Gaussian systematic effects.

Since data from both Gemini North and the Lick 3-m reject a constant, zero-polarization model with high confidence (Figure \ref{allmodels_zoom}), we assess the probability that data from the two telescopes are drawn from the same population. Specifically, we perform a $\chi^2$ analysis of Gemini North data against fits to Lick 3-m data and vice versa. We use three interpolating methods: splines, piecewise cubic Hermite interpolating polynomials (Matlab \texttt{pchip}), and modified Akima piecewise cubic Hermite interpolation (Matlab \texttt{makima}). These interpolants are of course unphysical, but they allow an assessment of whether data from the two telescopes are drawn from the same population agnostic of any physical, but potentially inaccurate, model. We choose not to employ a Kolmogorov-Smirnov test, as its nonparametric nature removes the clear dependence of orbital phase on measured polarization modulation.

Table \ref{conf} shows that no interpolating method may be rejected by any telescope's dataset, as confidences of rejection are well below $3 \sigma$. Thus, data from both telescopes and in both Stokes $q$ and $u$ are consistent with being drawn from the same population, and they all reject a zero polarization model with high confidence. We therefore conclude that both Gemini North and Lick 3-m have detected polarization modulation intrinsic to scattered light from HD 189733b. Table \ref{planetdata} lists best fit exoplanet parameters, where Stokes $\Delta I$ represents secondary eclipse depth, Stokes $\Delta (q,u,p)$ represent peak linear polarization values, and circular polarization Stokes $\Delta v$ will be discussed in Section \ref{circsec}. The negative sign in both peak Stokes $\Delta (q,u)$ indicates the longitude of the ascending node is $0^\circ < \Omega < 45^\circ$. Uncertainties in this table are given as $1\sigma$ unless otherwise specified.

\begin{deluxetable}{cccccccccc}
\tabletypesize{\normalsize}
\tablecaption{HD 189733b Model Rejection Confidence}
\tablewidth{0pt}
\tablehead{
\colhead{Model} & \colhead{8-m Data} & \colhead{3-m Data} & \colhead{8-m/3-m Data} }
\startdata
8-m spline        			& $-$ 	 	& $0.5\sigma$ 	 & $-$ \\
$\cdots$ \texttt{pchip}    	& $-$ 	 	& $0.2\sigma$ 	 & $-$ \\
$\cdots$ \texttt{makima}   	& $-$ 	 	& $0.1\sigma$ 	 & $-$ \\
\hline
3-m spline        			& $0.6\sigma$ 	 & $-$ 	 	& $-$ \\
$\cdots$ \texttt{pchip}    	& $0.7\sigma$ 	 & $-$ 	 	& $-$ \\
$\cdots$ \texttt{makima}   	& $0.4\sigma$ 	 & $-$ 	 	& $-$ \\
\hline
Zero $\Delta(q,u)$ 			& $7.2\sigma$ 	 & $2.9\sigma$ 	 & $7.2\sigma$   
\label{conf}
\enddata
\end{deluxetable}

\begin{deluxetable*}{cccccccccc}
\tabletypesize{\footnotesize}
\tablecaption{HD 189733 System Best Fit Parameters}
\tablewidth{0pt}
\tablehead{
\colhead{} & \colhead{Hubble \citep{Evans2013}} & \colhead{CHEOPS \citep{Krenn2023}} & \colhead{JWST \citep{Inglis2024}} & \colhead{POLISH2 (this work)}\\
				& ($\lambda = 390$ to 480 nm)	& ($\lambda = 350$ to 1100 nm)	& ($\lambda = 5.125$ to 11.75 $\mu$m)	& ($\lambda = 390$ to 475 nm)}
\startdata
Star + Planet \\
\hline
Stokes $\Delta I$ (ppm)			& $71 \pm 24$		& $24.7 \pm 4.5$		& 3 $143 \pm 6$		& $-$ \\
Stokes $\Delta q$ (ppm)			& $-$			& $-$				& $-$				& $-35.2 \pm 7.0$ \\
Stokes $\Delta u$ (ppm)			& $-$			& $-$				& $-$				& $-31 \pm 11$ \\
Stokes $\Delta v$ (ppm)			& $-$			& $-$				& $-$				& $67 \pm 19$ \\
Stokes $\Delta p$ (ppm)			& $-$			& $-$				& $-$				& $40.9 \pm 7.1$ \\
\hline
Planet \\
\hline
$p_\text{intrinsic}$ (\%)			& $-$			& $-$				& $-$				& $61.4 \pm 1.5$ \\
$A_g$						& $0.226 \pm 0.091$	& $0.076 \pm 0.016$		& $-$				& $> 0.26$ $(2\sigma)$ \\
SiO$_2$ $r_\text{eff}$ ($\mu$m)		& $-$		& $-$				& $0.0032^{+0.0127} _{-0.0025}$	& $0.038^{+0.047} _{-0.023}$ ($90\%$ conf.) \\
MgSiO$_3$ $r_\text{eff}$ ($\mu$m)		& $-$		& $-$				& $-$			& $0.035^{+0.047} _{-0.020}$ ($90\%$ conf.) \\
MnS $r_\text{eff}$ ($\mu$m)			& $-$		& $-$				& $-$			& $\sim 0.04$
\label{planetdata}
\enddata
\end{deluxetable*}

Magnetic activity in the host star HD 189733 has been implicated as the cause of statistically significant excursions in Stokes $q$ and $u$ measured by HIPPI and HIPPI-2 at the equatorially-mounted AAT 3.9-m telescope. Some excursions appear to be present from one night to the next \citep{Bailey2021}. However, we observe no such excursions in the orbital phase region where Lick 3-m and Gemini North POLISH2 observations overlap (Figure \ref{allmodels_zoom}), though we confine our analysis to this overlap region due to the presence of the outlier in Lick 3-m data outside this region as discussed above. Thus, unlike absolute photometry of the HD 189733 system, which is contaminated at the 1\% level by starspots modulated by the stellar rotation period \citep{Pont2007, Winn2007}, polarimetry measured over a seven-year period is clearly stable to at least three orders of magnitude deeper, to the 1 to 10 ppm level, when properly calibrated (Figure \ref{allmodels_zoom}). The prognosis for the study of exoplanets in polarized, scattered light is therefore more promising than previously reported \citep{Cotton2019xiBoo, Bailey2020, Bailey2021}.

\subsection{Scattering Models}

\begin{figure}
\centering
\includegraphics[width=0.47\textwidth]{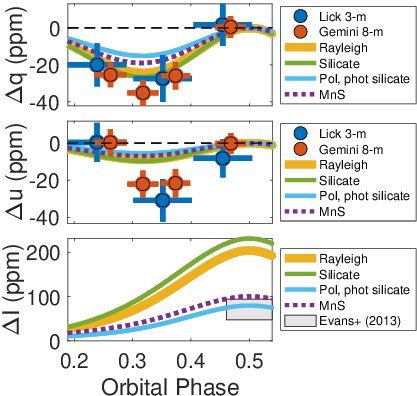}
\caption{Polarization and photometry of HD 189733b obtained in the Gemini North/Lick 3-m orbital phase overlap region. \textit{Top}: POLISH2 $B$ band Stokes $q$ measurements of HD 189733 obtained at both the Lick 3-m (blue points) and Gemini North (red points) and phased to the orbital period of the planet. Best fit Rayleigh (solid gold), silicate (SiO$_2$/MgSiO$_3$, solid green), and MnS (dashed purple) polarization models are shown. Shown in solid light blue is the best fit silicate model when combining POLISH2 polarimetry with Hubble photometry \citep{Evans2013}. \textit{Middle}: Same as at top but for Stokes $u$. \textit{Bottom}: Self-consistent photometric phase curves from the above polarization models in addition to a Hubble secondary eclipse measurement \citep{Evans2013}. Since secondary eclipse only occurs during a short interval centered at orbital phase 0.5, Hubble does not provide scattered light constraints during the rest of the planetary orbit.}
\label{data}
\end{figure}

The pure Rayleigh, SiO$_2$, MgSiO$_3$, and MnS models that best fit both Stokes $q$ and $u$ measurements simultaneously are illustrated in Figure \ref{data}. No model produces enough peak linear polarization to reproduce POLISH2 measurements. Since our polarization models also predict photometric phase curves, we include a Hubble $B$ band photometric constraint derived from a single, $3.0\sigma$ confidence HD 189733b secondary eclipse detection \citep{Evans2013}. This corresponds to a $B$ band geometric albedo of $A_g = 0.226 \pm 0.091$. While CHEOPS measured $A_g = 0.076 \pm 0.016$ for HD 189733b over thirteen secondary eclipses \citep{Krenn2023}, the wide, red CHEOPS bandpass (350 to 1100 nm) makes scaling to our $B$ band campaign difficult.

\begin{figure}
\centering
\includegraphics[width=0.47\textwidth]{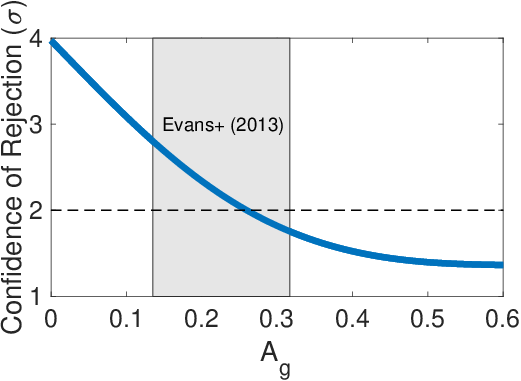}
\caption{$B$ band geometric albedo confidence limit from polarimetry data in the Gemini North orbital phase region (Figure \ref{allmodels_zoom}), which place a $2 \sigma$ lower limit to $B$ band geometric albedo of $A_g > 0.26$. The $1 \sigma$ range in $A_g$ from a Hubble secondary eclipse detection is shown in the grey region \citep{Evans2013}.}
\label{rejalb}
\end{figure}

For pure Rayleigh scattering, the model that fits our $B$ band polarization measurements best has single scattering and geometric albedos of 0.983 and 0.6, respectively, in this wavelength range. This implies a secondary eclipse depth of 172 ppm, which is $4.3 \sigma$ larger than that measured by Hubble \citep{Evans2013}. Given the significance of the polarization detection in the Gemini North orbital phase region (Figure \ref{allmodels_zoom}), we place a $2\sigma$ lower limit of $A_g > 0.26$ to the $B$ band geometric albedo of HD 189733b (Figure \ref{rejalb}). This $2\sigma$ lower limit from POLISH2 lies only $0.4 \sigma$ above the $B$ band Hubble geometric albedo of $A_g = 0.226 \pm 0.091$.

While unocculted starspots may mimic Rayleigh scattering in transmission spectroscopy \citep{McCullough2014}, they cannot do the same in linear polarimetry, as a realistic distribution of starspots on HD 189733 is modeled to cause time-variable, linear polarization contamination of $\Delta p \sim 1$ ppm or less \citep{Berdyugina2011, Kostogryz2015}. This is because stellar linear polarization is largest at the limb and rapidly vanishes toward the center of the stellar disk \citep{Chandrasekhar1946a, Chandrasekhar1946b, Kemp1983}. Starspots must lie at the limb to cause measurable change in stellar linear polarization, but their cross-sectional area vanishes at the limb. Therefore, by detecting linearly polarized, scattered light from HD 189733b near quarter phase, we have also invalidated unocculted starspots as the dominant cause of the blue optical Rayleigh-like slope in Hubble transmission spectroscopy. Note that the terms ``quarter phase" or ``dichotomy" describe a half-illuminated planet; the term ``quadrature" only applies to superior solar system bodies at a geometry where the observer, and not the object, is half-illuminated. Since exoplanets are always in interior geometries, they can never be observed at quadrature, and we caution the community against using this term.

\begin{figure}
\centering
\includegraphics[width=0.47\textwidth]{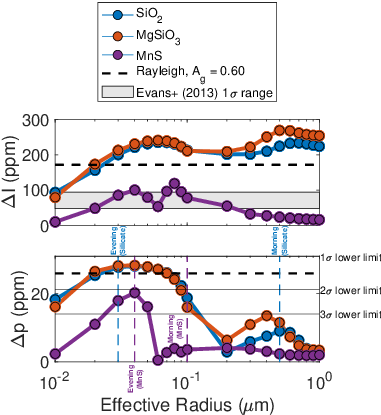}
\caption{\textit{Top}: Secondary eclipse depth (Stokes $\Delta I$) versus effective particle radius for SiO$_2$, MgSiO$_3$, and MnS particles. The pure Rayleigh secondary eclipse depth (dashed line) is calculated for geometric albedo $A_g = 0.60$. The Hubble secondary eclipse constraint is shown in the light grey box \citep{Evans2013}, which strongly favors MnS particles. \textit{Bottom}: Same as at top but for peak linear polarization (Stokes $\Delta p$), which strongly favors silicate particles. Lower limits to measured peak polarization are labeled at right. Best fit particle radii for data obtained at morning and evening terminators are shown (Section \ref{sec4}).}
\label{amplitudes}
\end{figure}

\begin{figure}
\centering
\includegraphics[width=0.47\textwidth]{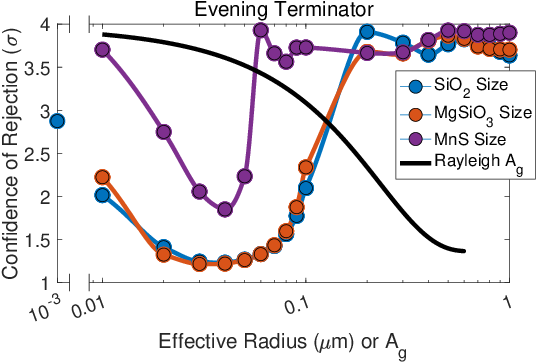}
\caption{Confidence of rejection of SiO$_2$, MgSiO$_3$, MnS, and pure Rayleigh models with various effective radii or geometric albedos based on $B$ band polarimetry data obtained near the HD 189733b evening terminator. The presence of 1 nm SiO$_2$ particles (the low end of sizes consistent with JWST secondary eclipse spectra, \citealt{Inglis2024}) is inconsistent with polarimetry with $2.9 \sigma$ confidence.}
\label{confrej}
\end{figure}

\begin{figure}
\centering
\includegraphics[width=0.47\textwidth]{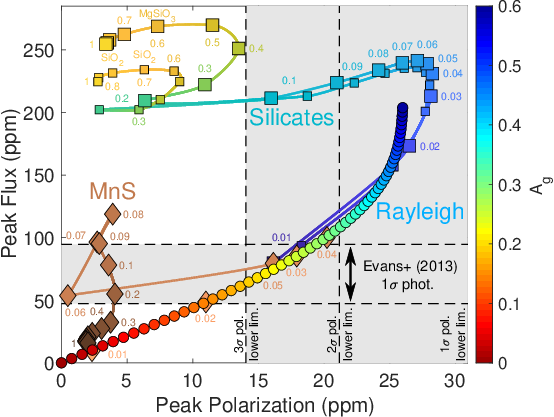}
\caption{Peak flux versus peak polarization of scattered light from HD 189733b dominated by pure Rayleigh, SiO$_2$, MgSiO$_3$, or MnS scatterers. Peak polarization and flux increase monotonically as single scattering albedo, and thereby geometric albedo $A_g$, is increased for Rayleigh scattering models. Silicate models deviate from pure Rayleigh models for effective particle radii $r_\text{eff} > 0.02$ $\mu$m, while MnS models appear Rayleigh-like for $r_\text{eff} < 0.05$ $\mu$m. SiO$_2$ and MgSiO$_3$ peak polarization diverges for $r_\text{eff} > 0.08$ $\mu$m, while they diverge in flux for $r_\text{eff} > 0.2$ $\mu$m. Large MgSiO$_3$ particles tend to be more reflective and polarized than SiO$_3$ of the same effective radius. Numbers attached to each model point indicate $r_\text{eff}$ of the scattering population in microns, and the color bar only applies to the Rayleigh scattering models. Measured $1\sigma$, $2\sigma$, and $3\sigma$ lower limits to measured linear polarization are shown.}
\label{photvspol}
\end{figure}

\begin{figure}
\centering
\includegraphics[width=0.47\textwidth]{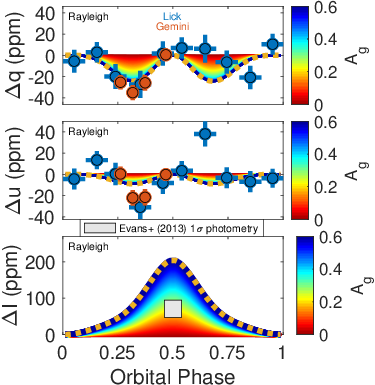}
\caption{Same as Figure \ref{data} but for pure Rayleigh scattering only. \textit{Top}: $B$ band Stokes $q$ linear polarization measurements of HD 189733 versus orbital phase compared to models containing pure Rayleigh scatterers. Blue and red data points indicate measurements obtained at the Lick 3-m and Gemini North, respectively. Peak polarization increases nonlinearly with increasing geometric albedo $A_g$. The best fit, pure Rayleigh model occurs for $A_g = 0.60$ (dashed gold curve), which is essentially the theoretical limit for a pure Rayleigh atmosphere. \textit{Middle}: Same as at top but for Stokes $u$. \textit{Bottom}: Planetary flux modulation $\Delta I$ versus orbital phase for pure Rayleigh models of varying geometric albedo. The \citet{Evans2013} Hubble secondary eclipse measurement, $A_g = 0.226 \pm 0.091$, is shown in the light grey box.}
\label{models_ray}
\end{figure}

\begin{figure}
\centering
\includegraphics[width=0.47\textwidth]{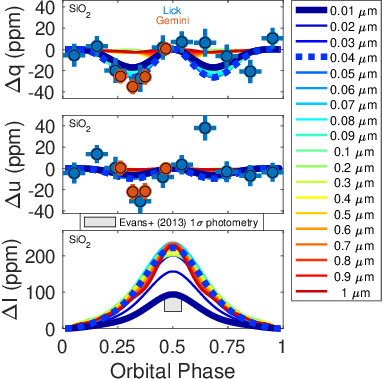}
\caption{Same as Figure \ref{models_ray} but for SiO$_2$ particles. While polarization measurements are best fit by $r_\text{eff} = 0.038^{+0.047} _{-0.023}$ $\mu$m particles (dashed blue curve at top and middle), Hubble secondary eclipse photometry \citep{Evans2013} is best fit by 0.01 $\mu$m particles (solid dark blue curve at bottom).}
\label{models_sio2}
\end{figure}

\begin{figure}
\centering
\includegraphics[width=0.47\textwidth]{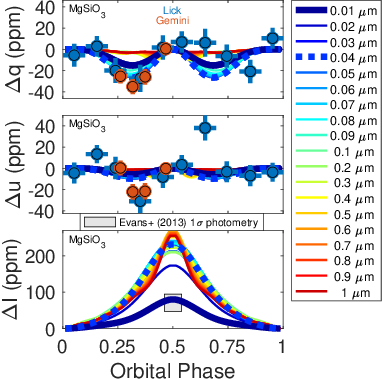}
\caption{Same as Figure \ref{models_ray} but for MgSiO$_3$ particles. While polarization measurements are best fit by $r_\text{eff} = 0.035^{+0.047} _{-0.020}$ $\mu$m particles (dashed blue curve at top and middle), Hubble secondary eclipse photometry \citep{Evans2013} is best fit by 0.01 $\mu$m particles (solid dark blue curve at bottom).}
\label{models_mgsio3}
\end{figure}

\begin{figure}
\centering
\includegraphics[width=0.47\textwidth]{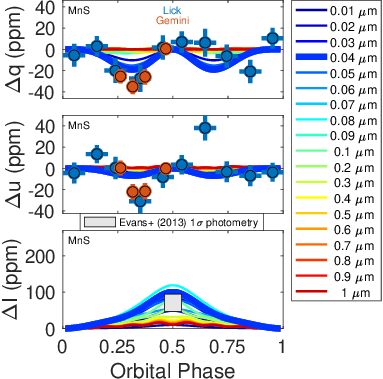}
\caption{Same as Figure \ref{models_ray} but for MnS particles. While polarization measurements are best fit by $r_\text{eff} \sim 0.04$ $\mu$m particles (solid blue curve at top and middle), the fit is worse than for silicates. Unlike for silicates, the MnS model that fits polarization measurements best also acceptably fits Hubble secondary eclipse photometry (solid blue curve at bottom) from \citet{Evans2013}.}
\label{models_mns}
\end{figure}


The best fit polarization phase curves for SiO$_2$, MgSiO$_3$, and MnS have effective particle radii of $r_\text{eff} = 0.038^{+0.047} _{-0.023}$, $0.035^{+0.047} _{-0.020}$ (90\% confidence) and $\sim 0.04$ $\mu$m, respectively (Table \ref{planetdata}, Figure \ref{amplitudes}, and Figure \ref{confrej}). However, such $0.02 < r_\text{eff} < 0.08$ $\mu$m silicate particles are too bright in backscattering to reproduce Hubble secondary eclipse photometry, as they generate secondary eclipse depths up to $6.4 \sigma$ larger than measured \citep{Evans2013}. Silicate particles backscatter so efficiently that $r_\text{eff}$ must be reduced to 0.01 $\mu$m in an attempt to simultaneously reproduce polarization and secondary eclipse depth. However, this further reduces modeled polarization significantly below measured values (Figure \ref{confrej}).

Retrievals of JWST secondary eclipse spectra strongly prefer SiO$_2$ to magnesium silicates such as MgSiO$_3$ \citep{Inglis2024}, but silicate $B$ band refractive indices are similar with and without magnesium. Thus, $B$ band polarimetry in this work is unable to distinguish between SiO$_2$ and MgSiO$_3$ within measurement accuracy. Absorption at 8.7 $\mu$m in JWST secondary eclipse spectra suggest the presence of $10^{-2.5 \pm 0.7}$ $\mu$m SiO$_2$ particles in HD 189733b's atmosphere \citep{Inglis2024}, which is generally smaller than the minimum particle radius of $0.01$ $\mu$m in our grid. However, modeling with the 3D Monte Carlo radiative transfer code POLARIS \citep{Reissl2016, Lietzow2021} suggests that a planet with 1 nm SiO$_2$ particles is only 0.54 times as polarized in $B$ band as one dominated by 0.01 $\mu$m particles (M. Michaelis, priv. comm.). Since even 0.01 $\mu$m particles have difficulty reproducing the large peak linear polarization measured by POLISH2 (Figures \ref{amplitudes} and \ref{confrej}), 1 nm particles are inconsistent with HD 187933b's polarization with $2.9 \sigma$ confidence (Figure \ref{confrej}).

The best fit MnS model, with $r_\text{eff} \sim 0.04$ $\mu$m, provides a secondary eclipse depth consistent with the literature \citep{Evans2013}, but it significantly underpredicts measured polarization (Figure \ref{amplitudes}). Indeed, MnS of all radii save 0.04 $\mu$m may be rejected with $> 2 \sigma$ confidence, while many silicate models are rejected with $< 2 \sigma$ confidence (Figure \ref{confrej}). MnS models are unable to reproduce JWST secondary eclipse spectroscopy more effectively than cloud-free models (J. Inglis, priv. comm.), which are less preferred over SiO$_2$ models with $\sim 6\sigma$ confidence \citep{Inglis2024}. Therefore, we reject MnS as a significant constituent of the HD 189733b atmosphere. Given the large amount of Hubble transmission spectroscopy, a Hubble secondary eclipse measurement, JWST secondary eclipse spectroscopy, and the $B$ band polarimetry in this work, we encourage the community to self-consistently model all of these data sources to pry HD 187933b's atmospheric secrets from its long-held grasp.

Figure \ref{photvspol} illustrates the complementary nature of photometry and polarimetry. For dark, pure Rayleigh scattering atmospheres with low geometric albedos (bottom left), both photometric and polarimetric phase curve peaks vary linearly with albedo. However, multiple scattering begins to be apparent in polarimetry for geometric albedos $A_g \ga 0.2$, and it acts to randomize the scattered electric field orientations. According to our models and \citet{Bailey2018}, this causes peak linear polarization to asymptote at $\sim 26$ ppm for HD 189733b.

The photopolarimetric signatures of SiO$_2$ and MgSiO$_3$ are quite different from that of pure Rayleigh scattering for $r_\text{eff} > 0.02$ $\mu$m. Silicate linear polarization peaks for $r_\text{eff} = 0.04$ $\mu$m particles but falls precipitously as effective radii are increased up to $r_\text{eff} = 0.2$ $\mu$m. Conversely, the photometric phase curve peak is nearly constant for $0.04 < r_\text{eff} < 0.2$ $\mu$m. Further increasing $r_\text{eff}$ causes a stereotypical Mie scattering oscillation, where polarization reaches a local maximum at $r_\text{eff} = 0.4$ $\mu$m. Peak polarization is weak but peak flux is large for large silicate particles. Photopolarimetric signatures of MnS particles, on the other hand, diverge from both pure Rayleigh and silicate particles for $r_\text{eff} > 0.04$ $\mu$m. Peak MnS polarization drops nearly to zero as effective radius is increased to 0.06 $\mu$m due to the large index of refraction of MnS. For $0.06 < r_\text{eff} < 1.0$ $\mu$m, peak polarization remains weak, while peak flux rises and falls. For $r_\text{eff} = 1.0$ $\mu$m, silicate and MnS particles harbor weak polarization, but an atmosphere composed of such particles would be very bright for silicates and very dark for MnS. Indeed, an atmosphere dominated by 1.0 $\mu$m silicate particles would have a secondary eclipse depth $\sim 15$ times larger than one dominated by MnS with the same particle radius distribution.

Polarization measurements are compared to the suite of pure Rayleigh, SiO$_2$, MgSiO$_3$, and MnS models in Figures \ref{models_ray} through \ref{models_mns}, respectively. All scatterers have significant difficulty simultaneously reproducing POLISH2 polarization and Hubble secondary eclipse measurements \citep{Evans2013}. Indeed, a high Rayleigh scattering geometric albedo or small silicate scatterers provide large polarization, but both generate a secondary eclipse depth that significantly exceeds the detection from Hubble. Conversely, a low enough Rayleigh scattering geometric albedo, or the smallest silicate particles in our grid, acceptably reproduce the Hubble secondary eclipse depth but fail to reproduce the large POLISH2 polarization. While high-index MnS particles may generate the low measured secondary eclipse depth, MnS has greater difficulty than silicates in explaining the large polarization measured in this work.

 While \citet{Lietzow2022} show linear polarization phase curves for a variety of species, these calculations were performed with 1 $\mu$m effective sizes. Given that Rayleigh-like particles are required to approach the large polarization of HD 189733b measured in this work, \citet{Lietzow2022} is not relevant here. Additionally, \citet{Bailey2018} show that the polarization of a planet dominated by 0.05 $\mu$m Fe and Al$_2$O$_3$ particles is 1/3 to 2/3 that of MgSiO$_3$ and Mg$_2$SiO$_4$, respectively, so we reject these species as the dominant cause of polarimetry measured in this work. Finally, \citet{Inglis2024} find no compelling evidence for the presence of cloud species other than SiO$_2$ in the HD 18973b atmosphere from JWST secondary eclipse spectroscopy.

\begin{figure}
\centering
\includegraphics[width=0.47\textwidth]{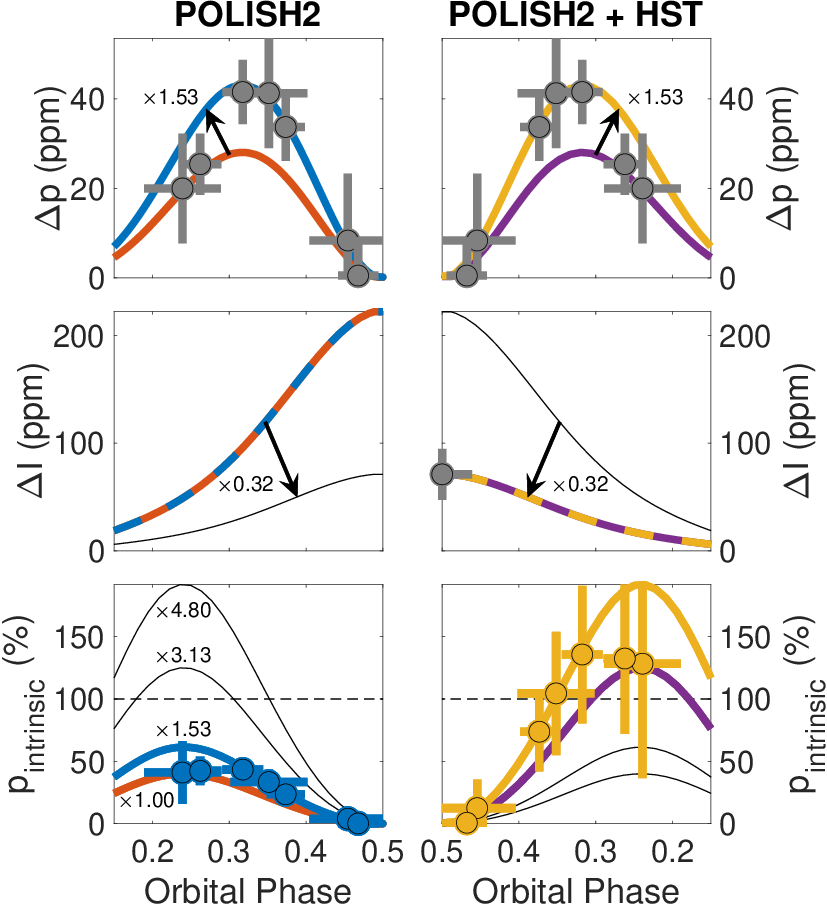}
\caption{Intrinsic $B$ band linear polarization of HD 189733b. \textit{Top}: POLISH2 polarization measurements (grey points) compared to the best-fit SiO$_2$ model (red and purple curves at left and right, respectively) as well as the model multiplied by a scaling factor of 1.53 to fit polarization measurements (blue and gold curves at left and right, respectively). Top left and right plots are identical, as the direction of the x-axis is reversed from left to right panels. \textit{Center}: Flux from the best-fit SiO$_2$ model (blue and red curves at left, thin black curve at right) compared to flux from the model multiplied by 0.32 (thin black curve at left, gold and purple curves at right) to fit a Hubble secondary eclipse measurement \citep[gold point at right]{Evans2013}. \textit{Bottom}: Intrinsic planetary linear polarization $p_\text{intrinsic} = \Delta p / \Delta I$ obtained by dividing polarization models in top panels (choice of two models blue = gold or red = purple) by flux models in center panels (choice of two models blue = red or gold = purple). This generates four intrinsic polarization combinations following blue, red, gold, and purple curves from top to bottom panels. Models that produce $p_\text{intrinsic} > 100\%$ are unphysical (horizontal, dashed line), which include those using \citet{Evans2013} secondary eclipse constraints. Blue (left panel) and gold data points (right panel) are shown, while red (left panel) and purple data points (right panel) are removed for clarity. Thin black curves at left and right mirror curves at right and left to guide the eye.}
\label{intrinsicpolmodels}
\end{figure}

\subsection{Comparison with Solar System Atmospheres}

\begin{figure}
\centering
\includegraphics[width=0.47\textwidth]{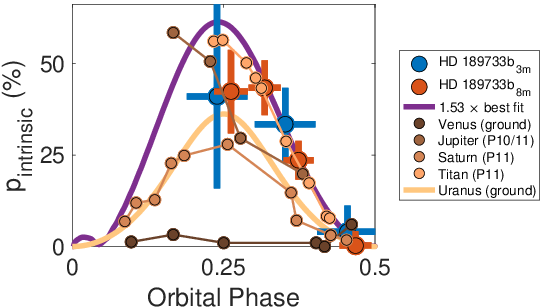}
\caption{Intrinsic $B$ band linear polarization of HD 189733b (blue and red data points) in the Gemini North orbital phase overlap region (Figure \ref{allmodels_zoom}). This is obtained by normalizing polarization measurements by the photometric phase curve of the best fit SiO$_2$ model (Figures \ref{data} and \ref{intrinsicpolmodels}, left center panel). The intrinsic polarization model (purple curve) is given by normalizing the best fit polarization phase curve, scaled by 1.53, to this photometric phase curve (blue curve at bottom left in Figure \ref{intrinsicpolmodels}). Also shown are blue optical measurements of solar system bodies measured from the ground \citep[Venus]{Lyot1929, Coffeen1969, Dollfus1970} and from Pioneer 10 and 11 \citep[Jupiter, Saturn, and Titan]{Tomasko1982, Smith1984, Tomasko1984}. Ground-based observations of Uranus, obtained at effective orbital phases $< 0.01$ \citep[Wiktorowicz et al., submitted]{Michalsky1977Uranus}, are consistent with Rayleigh scattering with $36.22\% \pm 0.89\%$ efficiency (gold curve). HD 189733b appears to have intrinsic linear polarization slightly larger than Jupiter and Titan. Phase angles of solar system observations are converted to HD 189733b's orbital phase.}
\label{intrinsicpol}
\end{figure}

Normalization of measured HD 189733b linear polarization $\Delta p$ by the flux curve of best fit models $\Delta I$ allows intrinsic planetary linear polarization $p_\text{intrinsic} = \Delta p / \Delta I$ to be estimated (Figure \ref{intrinsicpolmodels}). Polarization measurements from both Gemini North and the Lick 3-m in the Gemini North orbital phase overlap region (Figure \ref{allmodels_zoom}) are $1.534 \pm 0.038$ times larger than the most polarized SiO$_2$ model (Figure \ref{intrinsicpolmodels}, top panels). We propagate this model (red and purple curves in top left and right panels, respectively) and one scaled by a factor of 1.53 (blue and gold curves in top left and right panels, respectively) to calculate intrinsic planetary polarization.

Our most polarized SiO$_2$ model also predicts a flux curve, and it is not clear how scaling polarization by 1.53 would affect flux. We therefore normalize the above two polarization models (maximum polarization and further scaling) by the modeled flux curve of the maximum polarization model. The \citet{Evans2013} Hubble secondary eclipse measurement requires scaling of our flux curve by factor of $0.32 \pm 0.11$ (Figure \ref{intrinsicpolmodels}, center panels). In addition to normalizing the above two polarization models by our modeled flux curve, we also normalize these two polarization models by the flux curve scaled to fit Hubble secondary eclipse constraints.

\begin{deluxetable}{cccccccccc}
\tabletypesize{\normalsize}
\tablecaption{Peak Intrinsic $B$ Band Linear Polarization}
\tablewidth{0pt}
\tablehead{
 & \colhead{No $I$ Scaling} & \colhead{$I$ Scaling} }
\startdata
No $p$ Scaling & 40\%  	 		& 125(46)\% \\
$p$ Scaling    & \textbf{61.4(1.5)\%} 	& 192(71)\%   
\label{intrinsicpolmodelstable}
\enddata
\end{deluxetable}

\begin{deluxetable}{cccccccccc}
\tabletypesize{\normalsize}
\tablecaption{Peak Blue Optical Intrinsic Linear Polarization}
\tablewidth{0pt}
\tablehead{
\colhead{Object} & \colhead{$p_\text{intrinsic}$} & \colhead{References} }
\startdata
HD 189733b & $61.4(1.5)\%$	& This work \\
Jupiter    & $\sim 58\%$				& ST84\\
Titan      & $\sim 56\%$				& TS82  \\
Uranus     & 36.22(89)\% 				& MS77, W25b \\
Saturn     & $\sim 28\%$				& TD84 \\
Venus      & $\sim 6\%$				& L29, CG69, DC70    
\label{solarsys}
\enddata
\tablerefs{L29 \citep{Lyot1929}, CG69 \citep{Coffeen1969}, DC70 \citep{Dollfus1970}, MS77 \citep{Michalsky1977Uranus}, TS82 \citep{Tomasko1982}, ST84 \citep{Smith1984}, TD84 \citep{Tomasko1984}, W25b (Wiktorowicz et al., submitted) }
\end{deluxetable}

The combination of two polarization models (maximum polarization and scaling by 1.53 to account for polarization data) and two flux models (flux curve from the maximum polarization model and scaling by 0.32 to account for secondary eclipse data) results in four estimates of intrinsic planetary linear polarization (Figure \ref{intrinsicpolmodels}, bottom panels, and Table \ref{intrinsicpolmodelstable}, bold value preferred by data). The planet cannot deliver more net polarized flux than total flux, so $p_\text{intrinsic} < 100\%$. This calls the low Hubble secondary eclipse depth into question, as it forces intrinsic linear polarization of the planet to be $p_\text{intrinsic} \ga 100\%$. In contrast, polarization data normalized by our modeled flux curve identify intrinsic HD 189733b blue optical linear polarization similar to Jupiter and Titan. In the blue optical, HD 189733b is twice as intrinsically linearly polarized as Saturn and Uranus, and it is an order of magnitude more polarized than Venus (Figure \ref{intrinsicpol} and Table \ref{solarsys}).

As an interior planet, Venus harbors an extensive dataset at a large range of phase angles (angle between sun, object, and observer) and wavelengths of observation \citep{Lyot1929, Coffeen1969, Dollfus1970}. However, high phase angle observations of superior bodies such as Jupiter, Saturn, and Titan require space-based observatories due to limitations of Earth-based viewing geometry. Such datasets have been obtained during flybys of Pioneer 10 and 11 for Jupiter \citep{Smith1984} and of Pioneer 11 for both Saturn \citep{Tomasko1984} and Titan \citep{Tomasko1982}. Ground-based polarimetry of Uranus is measured to be $p \la 500$ ppm at phase angles $\la 3^\circ$ (effective orbital phase $< 0.01$), and its phase angle dependence is consistent with Rayleigh scattering with $p_\text{intrinsic} = 36.22\% \pm 0.89\%$ efficiency \citep[Wiktorowicz et al., submitted]{Michalsky1977Uranus}.

Comparison of these solar system bodies to HD 189733b shows the exoplanet to be strongly intrinsically polarized, where observed peak polarization is $p_\text{intrinsic} = 61.4 \pm 1.5\%$ (Tables \ref{intrinsicpolmodelstable} and \ref{solarsys}). Such large polarization is indicative of small, Rayleigh-like particles. The effect of particle size on photopolarization phase curves is illustrated by Venus, Titan, and Jupiter. Venus' large, $1.05 \pm 0.10$ $\mu$m sulfuric acid droplets generate weak polarization in the optical that peaks at low phase angle \citep{Hansen1974}, while Titan's small, $0.04 \pm 0.01$ $\mu$m haze monomers generate large polarization that peaks at quarter phase \citep{Tomasko2009}. Coincidentally, Titan's monomers are essentially the same size as the particles we detect in HD 189733b's atmosphere. However, these monomers are assembled in organic fractal aggregates named ``tholins" \citep{Sagan1979}, which are required to explain both large polarization (indicative of small particles) and large forward scattering brightness (indicative of large particles) given Titan’s haze composition \citep{Rages1983, West1991, WestSmith1991}. Interestingly, Jupiter also has evidence for fractal organic hazes given the same combination of large polarization and forward scattering \citep{West1991}.

\begin{figure}
\centering
\includegraphics[width=0.47\textwidth]{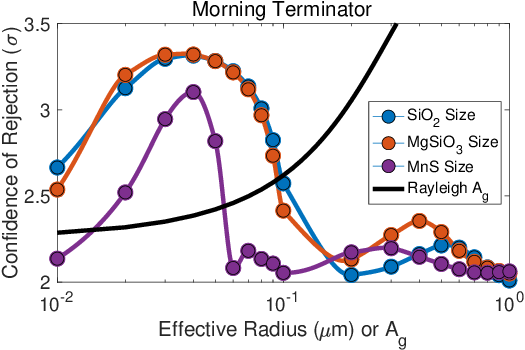}
\caption{Confidence of rejection of SiO$_2$, MgSiO$_3$, MnS, and pure Rayleigh models with various effective radii or geometric albedos based on $B$ band polarimetry data obtained near the HD 189733b morning terminator. This is in contrast to evening terminator data (Figure \ref{confrej}) and may suggest grain growth on the planet's nightside.}
\label{morning}
\end{figure}

\begin{figure*}
\centering
\includegraphics[width=0.9\textwidth]{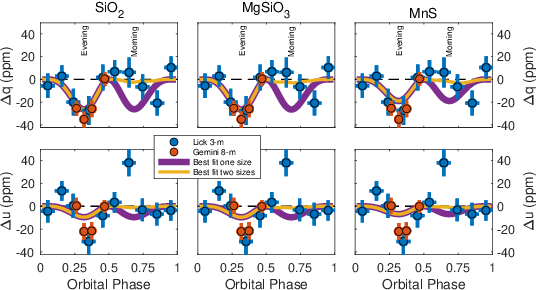}
\caption{\textit{Left}: Stokes $q$ (top) and $u$ (bottom) data along with best-fit SiO$_2$ models with one- (evening only, purple curve) and two-particle size distributions (evening and morning, gold curve), whose sizes are given in Table \ref{partgrowth}. One- and two-particle models overlap at the evening terminator, but they diverge at the morning terminator potentially due to grain growth via condensation on the nightside of the planet. \textit{Center}: Same as at left but for MgSiO$_3$. \textit{Right}: Same as at left but for MnS.}
\label{terminators}
\end{figure*}

Since HD 189733b's forward-scattered brightness may only be probed during mid-transit, when direct light from the host star presents a severe calibration challenge for photometry, assessment of the potentially non-spherical, fractal nature of HD 189733b's cloud particles remains elusive. Unfortunately, polarimetry is not likely to aid this assessment, as polarization signatures would be dominated by the small diameter of the aggregate's monomers. However, HD 189733b is likely too hot to enable organic tholins to exist in its atmosphere \citep{Gao2020}, which suggests that its atmosphere harbors spherical cloud particles as opposed to photochemical haze aggregates.

\subsection{Possible Particle Growth at Morning Terminator}\label{sec4}

Though the accuracy of Lick 3-m POLISH2 observations may be called into question for orbital phases outside the Gemini North campaign (Figure \ref{allmodels}), Lick 3-m Stokes $q$ measurements suggest a tantalizing, potential decrease in peak linear polarization across the orbital phase $\sim 0.7$ peak compared to the orbital phase $\sim 0.3$ peak (Figure \ref{allmodels}). Given HD 189733b's true obliquity of $\psi = 7^{+12^\circ}_{-4^\circ}$ \citep{Cegla2016}, the phase $\sim 0.7$ peak probes the morning terminator, while the phase $\sim 0.3$ peak probes the evening terminator. Since the polarization of silicate and MnS particles decreases for large particle radii (Figure \ref{photvspol}), we model separate particle populations at both terminators. We also model the $B$ band geometric albedo for pure Rayleigh scatterers at both terminators separately.

\begin{deluxetable*}{cccccccccc}
\tablecaption{Best Fit Particle Composition vs. Terminator}
\tablewidth{0pt}
\tablehead{
\colhead{Species} & \colhead{Morning Terminator} & \colhead{Evening Terminator} }
\startdata
SiO$_2$		& $r > 0.10$ $\mu$m (99\% conf.),	& $0.02 < r < 0.08$ $\mu$m (90\% conf.) \\
$\cdots$		& $r < 0.01$ $\mu$m (99\% conf.) \\
MgSiO$_3$	& $r > 0.09$ $\mu$m (99\% conf.),	& $0.02 < r < 0.08$ $\mu$m (90\% conf.) \\
$\cdots$		& $r < 0.01$ $\mu$m (99\% conf.) \\
MnS			& $r > 0.05$ $\mu$m (99\% conf.),	& $r \sim 0.04$ $\mu$m \\
$\cdots$		& $r < 0.02$ $\mu$m (99\% conf.) \\
Rayleigh		& $A_g < 0.09$ (99\% conf.)		& $A_g > 0.26$ ($2 \sigma$)
\label{partgrowth}
\enddata
\end{deluxetable*}

The two-particle model is consistent with the presence of particles with larger effective radii at the morning terminator than at the evening terminator, which could be caused by particle condensation on the night side of the planet and particle photoevaporation on the dayside (Figures \ref{morning} to \ref{terminators} and Table \ref{partgrowth}). Similar condensation of iron \citep[WASP-76b]{Ehrenreich2020} and potentially silicate particles \citep[LTT 9779b]{Coulombe2025} has been observed on other exoplanets. The two-albedo Rayleigh scattering model is consistent with a dark, $B$ band albedo $A_g < 0.09$ at the morning terminator (99\% confidence) and a bright, $A_g > 0.26$ albedo at the evening terminator ($2 \sigma$ confidence). Future polarimetric observations of HD 189733b centered on its morning terminator by a large, alt-az telescope would be sufficient to test this hypothesis.

\subsection{HD 189733 System Circular Polarization}\label{circsec}

We denote circular polarization from a localized, spatially resolved region of a planet $v_\text{region} = V_\text{region}/I_\text{region}$. When integrated across the planetary disk, this quantity becomes $v_\text{intrinsic} = V_\text{intrinsic}/I_\text{intrinsic}$. Finally, when the colossal, direct contaminating light from the host star is included, this becomes $v$. Spatially resolved circular polarization up to $v_\text{region} \sim 0.01\%$ has been detected on Venus \citep{Kemp1971b}, Jupiter \citep{Kemp1971a, Kemp1971b, Michalsky1974}, and Saturn \citep{Swedlund1972}. Earth's atmosphere has been observed to have $v_\text{region} \sim 0.1\%$ at twilight \citep{Angel1972}. Models may explain spatially resolved circular polarization from planets up to the $v_\text{region} \sim 0.1\%$ level due to multiple scattering \citep{Rossi2018}. As unpolarized starlight scatters in the atmosphere to generate linear polarization, subsequent scattering of this linearly polarized light may generate circular polarization with of order 1 part in 100 efficiency. Thus, given HD 189733b's peak intrinsic linear polarization of $p_\text{intrinsic} = 61.4(1.5)\%$ (Table \ref{solarsys}), its maximum plausible spatially resolved circular polarization is perhaps $v_\text{region} \sim 0.6\%$. However, the above measurements and models show that the sign of circular polarization changes from the planetary northern to southern hemisphere, so disk-integrated circular polarization of a transiting exoplanet is expected to be $v_\text{intrinsic} = 0$ absent asymmetric planetary features. Even if the disk-integrated planet were intrinsically $v_\text{intrinsic} = v_\text{region} = 0.1\%$ circularly polarized, contamination from the direct light of the host star would reduce system circular polarization to $v < 1$ ppm.

\begin{figure}
\centering
\includegraphics[width=0.47\textwidth]{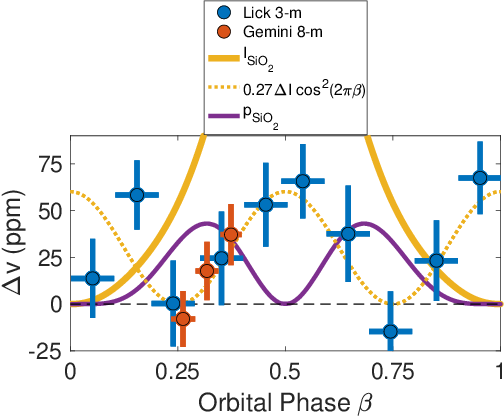}
\caption{$B$ band circular polarization (Stokes $v$) of the HD 189733 system measured at the Lick 3-m (blue points) and Gemini North (red points) as a function of HD 189733b's orbital phase. Both Lick 3-m and Gemini North measurements are strikingly consistent, which would require intrinsic planetary circular polarization to be $v_\text{intrinsic} = 27.0\% \pm 2.8\%$ (dashed gold curve). This is over two orders of magnitude larger than physically motivated models \citep{Rossi2018} and observations of solar system atmospheres. Near transit (orbital phases $< 0.25$ and $> 0.75$), the presence of circular polarization that is larger than the flux modulation from the planet (solid gold curve) precludes circular polarization originating from the planet. Therefore, another synchronously rotating mechanism appears to be present. Measured circular polarization is larger than modeled linear polarization modulation (purple curve).}
\label{planetcirc}
\end{figure}

Unexpectedly, therefore, we measure circular polarization from the HD 189733 system that appears phase-locked to the planet's orbit with peak $\Delta v = 67 \pm 19$ ppm and a 27 ppm square root of the weighted variance (Figure \ref{planetcirc}). Circular polarization modulation is significantly larger than the peak value of linear polarization from the planet ($\Delta p = 40.9 \pm 7.1$ ppm, Table \ref{planetdata}), and it is uncorrelated with the phase dependence of linear polarization modulation. The orbital phase dependence of this modulation is reproduced at both the Lick 3-m and Gemini North, and it follows the expected flux modulation of the planet near secondary eclipse (orbital phases 0.25 to 0.75), and it is uncorrelated with the phase dependence of linear polarization from the planet.

We identify and subtract a static, telescope circular polarization offset of $v_\text{offset} = 36$ ppm between the Lick 3-m and Gemini North zero points. This leaves a residual, weighted mean system plus telescope circular polarization of $v = 27.7 \pm 5.4$ ppm, which is difficult to explain in terms of either ISM circular polarization or unsubtracted telescope circular polarization. Similar to multiple scattering in planetary atmospheres \citep{Rossi2018}, the ISM may also convert linear to circular polarization with 1 part in 100 efficiency \citep{Kemp1972HD154445, Martin1974}. Indeed, ignoring the intrinsically circularly polarized, highly magnetic white dwarf Grw $+70^\circ8247$, we calculate the weighted mean ratio of circular to linear polarization in stars with detectable circular polarization \citep{Wiktorowicz2023} to be $v/p = 0.0122 \pm 0.0091$. However, this ratio is $v/p = 0.49 \pm 0.10$ for HD 189733, which is inconsistent with $v/p$ for the ISM with $4.7 \sigma$ confidence. Additionally, measured circular polarization modulation is inconsistent with zero with $4.9 \sigma$ confidence.

The planet would need to be $v_\text{intrinsic} = 27.0\% \pm 2.8\%$ circularly polarized to be consistent with measurements surrounding secondary eclipse (Figure \ref{planetcirc}, gold dashed curve). Unlike measured peak intrinsic linear polarization of $p_\text{intrinsic} = 61.4(1.5)\%$, which has corollaries in solar system atmospheres (Figure \ref{intrinsicpol}), such large, broadband circular polarization is exceedingly rare in nature. Indeed, the $v_\text{region} \sim 90\%$ circular polarization of scarab beetles \citep{McDonald2017} is only possible due to the presence of chiral molecules \citep{Michelson1911}. Circular polarization in the HD 189733 system is also observed near transit, at orbital phases $< 0.25$ and $> 0.75$, when the flux contribution from the planet is negligible from the best fit scattering models to linear polarization data. Since it is impossible for the planet to generate more circularly polarized flux than total photon flux ($v_\text{intrinsic} = V_\text{intrinsic} / I_\text{intrinsic} > 100\%$), the planet cannot be responsible for measured circular polarization near transit.

\begin{figure}
\centering
\includegraphics[width=0.47\textwidth]{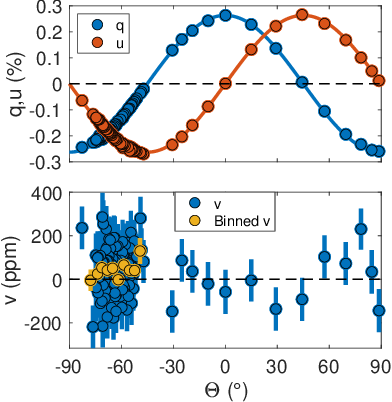}
\caption{\textit{Top}: Gemini POLISH HD 189733 Stokes $q$ (blue points) and $u$ (red points) measurements versus linear polarization orientation $\Theta$ from Figure \ref{gemrawdat} prior to telescope polarization subtraction. Solid lines indicate $q = q_{\text{max}} \cos 2 \Theta$ and $u = u_{\text{max}} \sin 2 \Theta$. \textit{Bottom}: Stokes $v$ measurements versus $\Theta$. No obvious sinusoidal trend is observed, suggesting residual instrumental crosstalk is negligible.}
\label{crosstalk}
\end{figure}

Calibration of circular polarization tends to be a difficult endeavor for conventional waveplate polarimeters. This is because optics upstream of the waveplate modulator impart unknown linear to circular polarization conversion into the instrument (``crosstalk"), and waveplates tend to have variations in retardance with wavelength. Since linear and circular polarization are rarely measured simultaneously, a complete picture of instrumental systematic effects is rarely obtained. However, POLISH2's photoelastic modulator pair measures linear and circular polarization simultaneously, so crosstalk is removed to the level of 1 part output circular polarization for 2,500 parts input linear polarization \citep{Wiktorowicz2023}. Due to the large linear telescope polarization measured at Gemini North (Figure \ref{gemrawdat}), whose amplitude is $46.0 \pm 2.8$ times larger than the interstellar linear polarization of the host star HD 189733 \citep{Wiktorowicz2023}, crosstalk would be identified by a correlation between measured circular polarization and telescope linear polarization. However, Figure \ref{crosstalk} shows no evidence for significant crosstalk in circular polarization measurements. Indeed, constant circular polarization as a function of linear polarization orientation $\Theta$ may only be rejected with $1.3 \sigma$ confidence. While the standard deviation of telescope polarization (Stokes $q$ and $u$) divided by median measurement uncertainty is 21, the same ratio for telescope Stokes $v$ is 1.1. Therefore, the variance in Stokes $v$ observations is consistent with measurement uncertainty and inconsistent with significant, residual crosstalk. This is consistent with POLISH2's ability to control crosstalk at the 1 part in 2,500 level, as telescope linear polarization of $p_\text{TP} = 0.2620 \pm 0.0034\% = 2620 \pm 34$ ppm \citep{Wiktorowicz2023} is only expected to generate residual circular polarization of $v_\text{TP} = 2620 / 2500 \sim 1$ ppm. Thus, measured peak circular polarization of $\Delta v = 67 \pm 19$ ppm (Figure \ref{planetcirc}) is inconsistent with a planetary or instrumental origin.

Individual sunspots have been observed to harbor broadband circular polarization up to $v_\text{region} \sim 0.1\%$ \citep{Illing1974a, Illing1974b, Illing1975, Kemp1982, Kemp1983sunspots}, especially in the blue optical. Both HD 189733 and HD 129333 are BY Dra variables, except HD 129333's starspot filling fraction is $\sim 6\%$ \citep{Dorren1994} while HD 189733's is 1\% \citep{Pont2007, Winn2007}. HD 129333 also has evidence for significantly larger circular polarization variability, in the blue optical, than we detect in HD 189733 \citep{Elias1990}. In a roughly 380 to 460 nm bandpass (similar to POLISH2's 390 to 475 nm $B$ bandpass), disk-integrated variability of HD 129333 is $\Delta v_\text{intrinsic} \sim 0.56\%$ from the square root of the weighted variance of measurements in \citet{Elias1990}. Constant blue optical circular polarization of HD 129333 may be rejected with $7.3 \sigma$ confidence, which highlights the variable nature of this intrinsic, $B$ band stellar circular polarization. In a roughly 500 to 580 nm bandpass, however, variability of HD 129333 drops to $\Delta v_\text{intrinsic} \sim 0.04\%$, and constant circular polarization may only be rejected with $0.9\sigma$ confidence. If circular polarimetric variance in BY Dra variables is linearly dependent on spot filling fraction, scaling HD 129333's circular polarimetric variability to the 1\% spot filling fraction of HD 189733 should result in $\Delta v_\text{intrinsic} = 0.56\% \times 1\% / 6 \% \sim 0.09\% \sim 900$ ppm for HD 189733. This is an order of magnitude larger variability than we measure for HD 189733 (peak $\Delta v = 67 \pm 19$ ppm and 28 ppm square root of the weighted variance).

Therefore, stellar activity appears to be a plausible mechanism to explain HD 189733's measured circular polarization modulation. Though our Lick 3-m campaign evenly samples orbital phase, it is striking that POLISH2 observations suggest stable patterns of stellar activity that generate circular polarization both near transit and secondary eclipse (Figure \ref{planetcirc}). This suggests star-planet interaction, stable over at least the seven years of POLISH2 observations, both at the sub-planetary point on the star and its antipode. Reports of star-planet interaction in the HD 189733 system are inconclusive, partly due to observations preferentially performed near transit or secondary eclipse \citep{Route2019a}. We caution that detailed modeling is required to assess the plausibility of stable, circularly polarized features on the HD 189733 disk that are phase-locked to the orbital period of the planet. We predict that the strongly wavelength dependent nature of starspot circular polarization \citep{Kemp1983sunspots, Elias1990} will be evident in future multiband measurements of HD 189733 with PHALANX.

\section{Extension to Known Exoplanets}

\begin{figure*}
\centering
\includegraphics[width=0.75\textwidth]{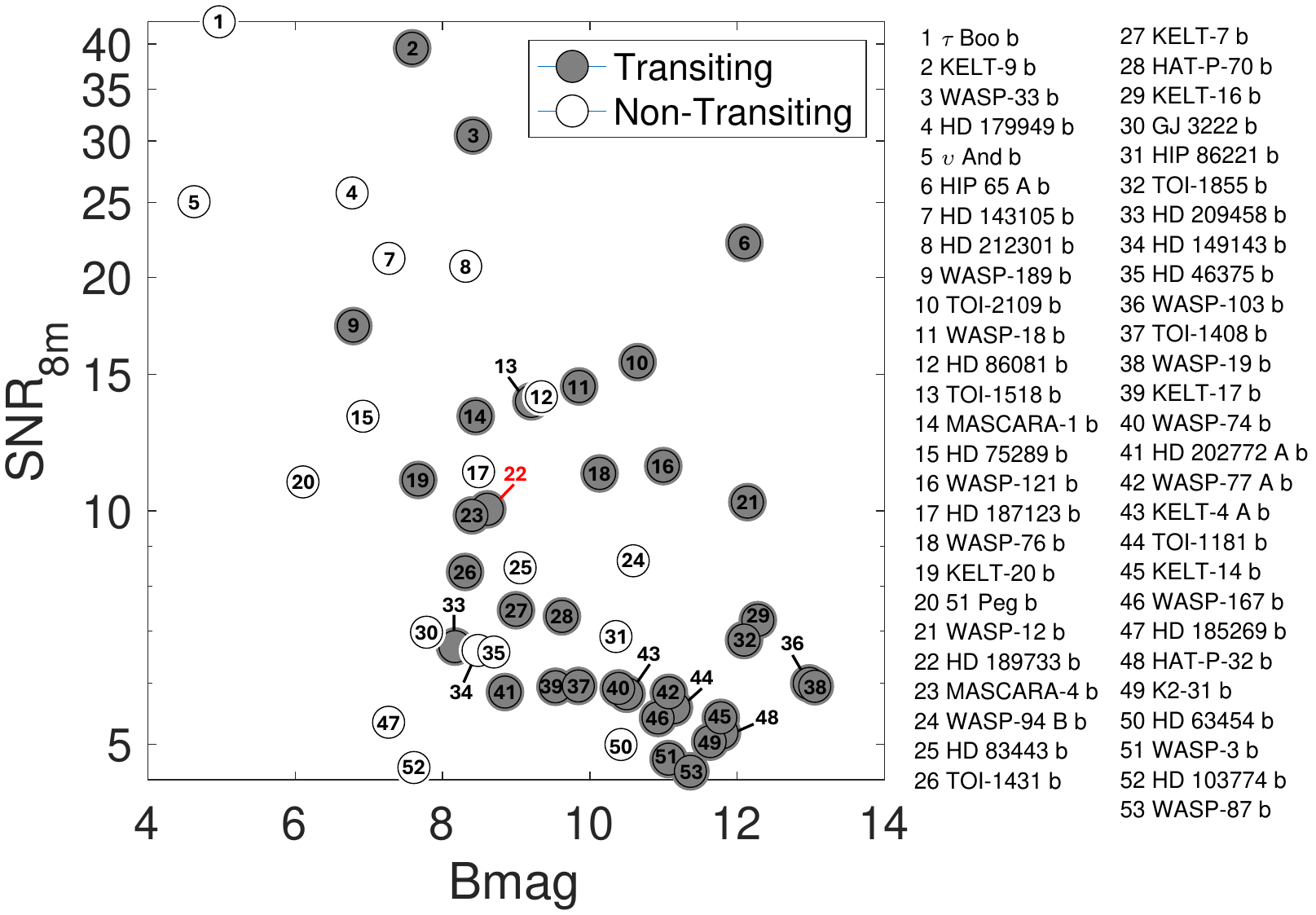}
\caption{Estimated SNR of detection of the known transiting (dark circles) and non-transiting (open circles) exoplanets vs. $B$mag assuming HD 189733-like atmospheres and one full orbit of observation at Gemini North (Table \ref{allplanets}). We model 21 exoplanets to be detectable with greater confidence than the HD 189733b detection in this work (red annotation). Exoplanet parameters are obtained from the NASA Exoplanet Archive.}
\label{allplanetsfig}
\end{figure*}

Clearly, self-calibration of telescope polarization at alt-az telescopes enables a massive increase in polarization accuracy for exoplanet characterization (Figure \ref{allmodels}). We model the expected SNR for planetary scattered light polarization at various alt-az telescopes by generating polarization phase curves for the exoplanets in the NASA Exoplanet Archive \citep{NASAExArch} and sampling them based on the noise statistics of our Gemini North data. Table \ref{allplanets} (sorted by SNR) and Figure \ref{allplanetsfig} show the exoplanets in the NASA Exoplanet Archive estimated to be detected with at least $5\sigma$ confidence in one week of observations with POLISH2 at Gemini North. Expected SNR obtained at 6.5-m and 1-m alt-az telescopes is also listed. Table \ref{subjov} (sorted by $M, M \sin i$) lists the expected yield of sub-jovians ($M, M \sin i < 115 \, M_\oplus$) detected with at least $3\sigma$ confidence in one week of observations at 30-m class telescopes. We choose this mass value because the demarcation between a power law $M$-$R$ trend for sub-jovians and $M$-$R$ independence for jovians occurs at $115 \, M_\oplus \pm 19 \, M_\oplus$ \citep{Edmondson2023}. Finally, Table \ref{eccen} (sorted by $T_\text{eq}$) shows $T_\text{eq} \leq 1000 \, K$ exoplanets expected to be detected with least $3\sigma$ confidence after one week of time at 30-m class telescopes.

Generally, since POLISH2 is photon noise limited \citep{WiktorowiczLaughlin2014}, telescope aperture $D$, planetary radius $R$, semimajor axis $a$ (for circular orbits), stellar brightness $B\text{mag}$, and number $n$ of orbits observed, relative to the HD 189733b detection in this work, affect SNR according to the following:

\begin{eqnarray}
\label{snrscaling}
\text{SNR} & = & 7.2 \left(\frac{D}{8\text{ m}}\right) \left(\frac{R / 1.138\text{ }R_J}{a / 0.031\text{ au}}\right)^2 \\
\nonumber & \times& \sqrt{\left(\frac{n}{0.5\text{ orbits}}\right) 10^{0.4 (8.578 - B\text{mag})}}.
\end{eqnarray}

\begin{deluxetable*}{lrcccccccccccc}
\tabletypesize{\scriptsize}
\tablecaption{Estimated SNR for Known Exoplanets}
\tablewidth{0pt}
\tablehead{
\colhead{Planet} & \colhead{Dec} & \colhead{$M$, $M \sin i$} & \colhead{$R$} & \colhead{$a$}	& \colhead{$e$} & \colhead{$\omega$} & \colhead{$B$mag} & \colhead{$T_*$} & \colhead{$R_*$} & \colhead{$T_\text{eq}$}& \colhead{SNR$_\text{8m}$} & \colhead{SNR$_\text{6.5m}$} & \colhead{SNR$_\text{1m}$} \\
& ($^\circ$) & ($M_J$) & ($R_J$) & (au) & & ($^\circ$) & & ($K$) & ($R_\odot$) & ($K$) & (1 week) & (1 week) & (1 month) }
\startdata
$\tau$ Boo b                  	& +17  	&  4.3 	& $\sim1.3$	& 0.049	& 0.010	&  88	&  5.0	&  6460	& 1.4	& 1690	& 43	& 35	& 10 \\
KELT-9 b\tablenotemark{*}     	& +40  	&  2.9 	& 1.9	& 0.035	& 0.000	&  90	&  7.6	& 10170	& 2.4	& 4050	& 40	& 32	&  8 \\
WASP-33 b\tablenotemark{*}    	& +38  	&  2.1 	& 1.6	& 0.025	& 0.000	&  90	&  8.4	&  7430	& 1.4	& 2750	& 30	& 24	&  8 \\
HD 179949 b                   	& $-$24	&  0.9 	& $\sim1.3$	& 0.044	& 0.022	& 190	&  6.8	&  6180	& 1.3	& 1590	& 26	& 21	&  5 \\
$\upsilon$ And b              	& +41  	&  0.7 	& $\sim1.3$	& 0.059	& 0.021	& 313	&  4.6	&  6140	& 1.6	& 1550	& 25	& 24	&  7 \\
HIP 65 A b\tablenotemark{*}   	& $-$55	&  3.2 	& 2.0	& 0.018	& 0.000	&  90	& 12.1	&  4590	& 0.7	& 1410	& 22	& 18	&  5 \\
HD 143105 b                   	& +69  	&  1.2 	& $\sim1.2$	& 0.038	& 0.035	&  90	&  7.3	&  6380	& $-$	&  $-$	& 21	& 17	&  5 \\
HD 212301 b                   	& $-$78	&  0.5 	& $\sim1.4$	& 0.033	& 0.000	&  90	&  8.3	&  6250	& 1.2	& 1790	& 21	& 17	&  5 \\
\hline
WASP-189 b\tablenotemark{*}   	& $-$3 	&  2.0 	& 1.6	& 0.051	& 0.000	&  90	&  6.8	&  8000	& 2.4	& 3350	& 17	& 14	&  4 \\
TOI-2109 b\tablenotemark{*}   	& +17  	&  5.0 	& 1.3	& 0.018	& 0.000	&  90	& 10.6	&  6540	& 1.7	& 3650	& 16	& 12	&  4 \\
WASP-18 b\tablenotemark{*}    	& $-$46	& 10.3 	& 1.2	& 0.020	& 0.008	& 157	&  9.9	&  6430	& 1.3	& 2410	& 14	& 12	&  3 \\
HD 86081 b                    	& $-$4 	&  1.6 	& $\sim1.4$	& 0.035	& 0.010	&  65	&  9.3	&  6020	& 1.5	& 1930	& 14	& 11	&  3 \\
TOI-1518 b\tablenotemark{*}   	& +67  	&  2.3 	& 1.9	& 0.039	& 0.005	&  90	&  9.2	&  7300	& 1.9	& 2490	& 14	& 11	&  3 \\
MASCARA-1 b\tablenotemark{*}  	& +11  	&  3.7 	& 1.6	& 0.040	& 0.000	& 344	&  8.4	&  7520	& 2.1	& 2590	& 13	& 11	&  3 \\
HD 75289 b                    	& $-$42	&  0.5 	& $\sim1.2$	& 0.048	& 0.015	&  50	&  6.9	&  6120	& 1.2	& 1490	& 13	& 11	&  3 \\
WASP-121 b\tablenotemark{*}   	& $-$39	&  1.2 	& 1.8	& 0.026	& 0.000	&  10	& 11.0	&  6460	& 1.5	& 2710	& 11	&  9	&  3 \\
HD 187123 b                   	& +34  	&  0.5 	& $\sim1.2$	& 0.042	& 0.007	&  44	&  8.5	&  5820	& 1.2	& 1470	& 11	&  9	&  2 \\
WASP-76 b\tablenotemark{*}    	& +3   	&  0.9 	& 1.8	& 0.033	& 0.016	&  62	& 10.1	&  6320	& 1.7	& 2200	& 11	&  9	&  3 \\
KELT-20 b\tablenotemark{*}    	& +31  	&  $-$ 	& 1.8	& 0.054	& 0.000	&  90	&  7.7	&  8940	& 1.6	& 2260	& 11	&  9	&  3 \\
51 Peg b                      	& +21  	&  0.5 	& $\sim1.1$	& 0.052	& 0.005	&   6	&  6.1	&  5760	& 1.2	& 1320	& 11	&  9	&  3 \\
WASP-12 b\tablenotemark{*}    	& +30  	&  1.5 	& 1.9	& 0.023	& 0.040	& 286	& 12.1	&  6280	& 1.7	& 2550	& 10	&  8	&  2 \\
\textbf{HD 189733 b}\tablenotemark{*}  	& +23  	&  1.2 	& 1.1	& 0.031	& 0.002	&  20	&  8.6	&  5050	& 0.8	& 1200	& 10	&  8	&  2 \\
MASCARA-4 b\tablenotemark{*}  	& $-$66	&  3.1 	& 1.5	& 0.047	& 0.000	&  90	&  8.4	&  7800	& 1.9	& 2100	& 10	&  8	&  2 \\
WASP-94 B b                   	& $-$34	&  0.6 	& $\sim1.4$	& 0.034	& 0.000	&  90	& 10.6	&  6040	& 1.4	& 1850	&  9	&  7	&  2 \\
HD 83443 b                    	& $-$43	&  0.4 	& $\sim1.1$	& 0.041	& 0.013	& 130	&  9.1	&  5440	& 1.0	& 1280	&  8	&  7	&  2 \\
TOI-1431 b\tablenotemark{*}   	& +56  	&  3.1 	& 1.5	& 0.046	& 0.002	& 108	&  8.3	&  7690	& 1.9	& 2370	&  8	&  7	&  2 \\
KELT-7 b\tablenotemark{*}     	& +33  	&  1.3 	& 1.5	& 0.044	& 0.000	&  90	&  9.0	&  6790	& 1.7	& 2050	&  7	&  6	&  2 \\
HAT-P-70 b\tablenotemark{*}   	& +10  	&  6.8 	& 1.9	& 0.047	& 0.000	&  90	&  9.6	&  8450	& 1.9	& 2560	&  7	&  6	&  2 \\
KELT-16 b\tablenotemark{*}    	& +32  	&  2.8 	& 1.4	& 0.020	& 0.000	&  90	& 12.3	&  6240	& 1.4	& 2450	&  7	&  6	&  1 \\
\textbf{GJ 3222 b}                     	& $-$40	&  \textbf{0.04}	& $\sim0.3$	& 0.091	& 0.929	& 291	&  7.8	&   $-$	& $-$	&  $-$	&  7	&  6	&  1 \\
HIP 86221 b                   	& +28  	&  0.7 	& $\sim1.2$	& 0.031	& 0.086	& 209	& 10.4	&   $-$	& $-$	&  $-$	&  7	&  6	&  2 \\
TOI-1855 b\tablenotemark{*}   	& +18  	&  1.1 	& 1.6	& 0.024	& 0.033	& 260	& 12.1	&  5360	& 1.0	& 1700	&  7	&  5	&  2 \\
HD 209458 b\tablenotemark{*}  	& +19  	&  0.7 	& 1.3	& 0.047	& 0.005	&   0	&  8.2	&  6070	& 1.2	& 1450	&  7	&  5	&  2 \\
HD 149143 b                   	& +2   	&  1.5 	& $\sim1.3$	& 0.053	& 0.017	& 184	&  8.5	&  5940	& 1.6	& 1560	&  7	&  5	&  2 \\
\textbf{HD 46375 b}                    	& +5   	&  \textbf{0.2} 	& $\sim0.9$	& 0.040	& 0.059	&  39	&  8.7	&  5260	& 0.9	&  $-$	&  7	&  5	&  2 \\
WASP-103 b\tablenotemark{*}   	& +7   	&  1.5 	& 1.6	& 0.020	& 0.075	&  90	& 13.0	&  6110	& 1.4	& 2500	&  6	&  5	&  1 \\
TOI-1408 b\tablenotemark{*}   	& +73  	&  1.9 	& 2.2	& 0.058	& 0.002	& 274	&  9.8	&  6210	& 1.4	& 1460	&  6	&  5	&  1 \\
WASP-19 b\tablenotemark{*}    	& $-$46	&  1.1 	& 1.4	& 0.016	& 0.005	&  65	& 13.1	&  5530	& 1.0	& 2080	&  6	&  5	&  1 \\
KELT-17 b\tablenotemark{*}    	& +14  	&  1.3 	& 1.5	& 0.049	& 0.000	&  90	&  9.5	&  7450	& 1.6	& 2090	&  6	&  5	&  1 \\
WASP-74 b\tablenotemark{*}    	& $-$1 	&  0.8 	& 1.4	& 0.035	& 0.015	&  90	& 10.4	&  5980	& 1.6	& 1920	&  6	&  5	&  1 \\
HD 202772 A b\tablenotemark{*}	& $-$27	&  1.0 	& 1.5	& 0.052	& 0.038	&  98	&  8.8	&  6270	& 2.6	& 2130	&  6	&  5	&  1 \\
WASP-77 A b\tablenotemark{*}  	& $-$7 	&  1.7 	& 1.2	& 0.024	& 0.005	& 194	& 11.1	&  5540	& 0.9	& 1690	&  6	&  5	&  1 \\
KELT-4 A b\tablenotemark{*}   	& +26  	&  0.9 	& 1.7	& 0.043	& 0.000	&  90	& 10.5	&  6210	& 1.6	& 1820	&  6	&  5	&  2 \\
\hline
TOI-1181 b\tablenotemark{*}   	& +64  	&  1.2 	& 1.4	& 0.036	& 0.000	&  90	& 11.1	&  6050	& 1.9	& 1940	&  6	&  4	&  1 \\
KELT-14 b\tablenotemark{*}    	& $-$42	&  1.3 	& 1.7	& 0.030	& 0.000	&  90	& 11.8	&  5770	& 1.5	& 1940	&  5	&  4	&  1 \\
WASP-167 b\tablenotemark{*}   	& $-$36	&  8.0 	& 1.6	& 0.036	& 0.000	&  90	& 10.9	&  7000	& 1.8	& 2330	&  5	&  4	&  1 \\
HD 185269 b                   	& +28  	&  1.0 	& $\sim1.2$	& 0.077	& 0.239	& 176	&  7.3	&  6040	& 2.0	& 1480	&  5	&  4	&  1 \\
HAT-P-32 b\tablenotemark{*}   	& +47  	&  0.7 	& 1.8	& 0.034	& 0.045	&  52	& 11.8	&  6180	& 1.2	& 1830	&  5	&  4	&  1 \\
K2-31 b\tablenotemark{*}      	& $-$24	&  1.8 	& 1.2	& 0.023	& 0.000	&  90	& 11.6	&  5350	& 0.9	& 1610	&  5	&  4	&  1 \\
HD 63454 b                    	& $-$78	&  0.3 	& $\sim1.2$	& 0.038	& 0.000	&  87	& 10.4	&  4820	& 0.8	&  $-$	&  5	&  4	&  1 \\
WASP-3 b\tablenotemark{*}     	& +36  	&  1.9 	& 1.4	& 0.032	& 0.003	&  90	& 11.1	&  6390	& 1.4	& 2010	&  5	&  4	&  1 \\
HD 103774 b                   	& $-$12	&  0.4 	& $\sim1.1$	& 0.070	& 0.090	& 318	&  7.6	&  6490	& $-$	&  $-$	&  5	&  4	&  1 \\
WASP-87 b\tablenotemark{*}    	& $-$53	&  2.2 	& 1.4	& 0.029	& 0.000	&  90	& 11.4	&  6450	& 1.6	& 2310	&  5	&  4	&  1   
\label{allplanets}
\enddata
\tablenotetext{*}{Transiting planet}
\end{deluxetable*}

\begin{deluxetable*}{lrcccccccccccc}
\tabletypesize{\scriptsize}
\tablecaption{Estimated SNR for $M, M \sin i < 115 \, M_\oplus$ Exoplanets}
\tablewidth{0pt}
\tablehead{
\colhead{Planet} & \colhead{Dec} & \colhead{$M, M \sin i$} & \colhead{$R$} & \colhead{$a$}	& \colhead{$e$} & \colhead{$\omega$} & \colhead{$B$mag} & \colhead{$T_*$} & \colhead{$R_*$} & \colhead{$T_\text{eq}$} & \colhead{SNR$_\text{30m}$} & \colhead{SNR$_\text{8m}$} \\
& ($^\circ$) & ($M_\oplus$) & ($R_\oplus$) & (au) & & ($^\circ$) & & ($K$) & ($R_\odot$) & ($K$) & (1 week) & (1 week) }
\startdata
HD 155918 b                 	& $-$75	&   8	&  $\sim3$	& 0.030	& 0.19	&  92	&  7.6	&  $-$	& $-$	&  $-$	&  5	& 2 \\
55 Cnc e\tablenotemark{*}   	& +28  	&   8	&  2	& 0.016	& 0.05	& 123	&  6.8	& 5200	& 0.9	& 1960	&  4	& 2 \\
HD 134060 b                 	& $-$61	&  10	&  $\sim3$	& 0.044	& 0.45	& 262	&  6.9	& 5970	& $-$	&  $-$	&  5	& 2 \\
HD 12235 b                  	& +3   	&  10	&  $\sim4$	& 0.054	& 0.38	& 279	&  6.5	&  $-$	& $-$	&  $-$	&  3	& 2 \\
GJ 3222 b\tablenotemark{a}                   	& $-$40	&  11	&  $\sim4$	& 0.091	& 0.93	& 291	&  7.8	&  $-$	& $-$	&  $-$	& 25	& 7 \\
HD 24085 b                  	& $-$70	&  12	&  $\sim4$	& 0.034	& 0.22	&   9	&  8.2	& 6030	& $-$	&  $-$	&  7	& 2 \\
\hline
HD 219828 b                 	& +19  	&  20	&  $\sim5$	& 0.050	& 0.08	& 240	&  8.7	& 5880	& 1.6	&  $-$	&  4	& 1 \\
HD 47186 b                  	& $-$28	&  23	&  $\sim5$	& 0.050	& 0.04	&  59	&  8.3	& 5660	& 1.1	&  $-$	&  4	& 1 \\
LTT 9779 b\tablenotemark{*} 	& $-$38	&  29	&  5	& 0.017	& 0.00	&  90	& 10.6	& 5440	& 0.9	& 1980	&  6	& 2 \\
HD 115404 A b               	& +17  	&  31	&  $\sim6$	& 0.088	& 0.23	& 259	&  7.5	&  $-$	& $-$	&  $-$	&  3	& 1 \\
HD 49674 b                  	& +41  	&  35	&  $\sim7$	& 0.058	& 0.09	&  17	&  8.8	& 5610	& 1.0	&  $-$	&  4	& 1 \\
WASP-127 b\tablenotemark{*} 	& $-$4 	&  55	& 15	& 0.051	& 0.00	&  90	& 10.8	& 5620	& 1.4	& 1400	&  7	& 2 \\
WASP-18 c                   	& $-$46	&  55	&  $\sim9$	& 0.035	& 0.01	& 272	&  9.9	& 6400	& 1.2	&  $-$	& 12	& 3 \\
KELT-11 b\tablenotemark{*}  	& $-$9 	&  63	& 15	& 0.062	& 0.05	& 116	&  8.8	& 5330	& 2.7	& 1710	& 11	& 3 \\
HD 76700 b                  	& $-$67	&  66	& $\sim10$	& 0.050	& 0.03	&  90	&  8.8	& 5640	& 1.4	&  $-$	& 11	& 3 \\
TOI-1273 b\tablenotemark{*} 	& +58  	&  71	& 11	& 0.055	& 0.06	& 110	& 11.6	& 5690	& 1.1	& 1210	&  3	& 1 \\
HD 46375 b                  	& +5   	&  72	& $\sim10$	& 0.040	& 0.06	&  39	&  8.7	& 5260	& 0.9	&  $-$	& 24	& 7 \\
HD 168746 b                 	& $-$12	&  75	& $\sim10$	& 0.065	& 0.09	&  42	&  8.7	& 5580	& 1.1	&  $-$	&  8	& 2 \\
TOI-1199 b\tablenotemark{*} 	& +61  	&  76	& 10	& 0.050	& 0.03	&  57	& 11.7	& 5710	& 1.4	& 1490	&  3	& 1 \\
WASP-69 b\tablenotemark{*}  	& $-$5 	&  82	& 12	& 0.045	& 0.06	&  90	& 10.9	& 4700	& 0.8	&  960	&  5	& 1 \\
HD 108147 b\tablenotemark{a}                 	& $-$64	&  83	& $\sim11$	& 0.102	& 0.50	& 319	&  7.5	& 6260	& 1.2	&  $-$	& 13	& 4 \\
WASP-131 b\tablenotemark{*} 	& $-$31	&  86	& 13	& 0.061	& 0.00	&  90	& 10.6	& 6030	& 1.5	& 1460	&  4	& 1 \\
HD 109749 b                 	& $-$41	&  89	& $\sim11$	& 0.061	& 0.01	& 110	&  8.9	& 5870	& 1.3	&  $-$	&  9	& 3 \\
HD 88133 b                  	& +18  	&  91	& $\sim12$	& 0.047	& 0.05	& 202	&  8.8	& 5440	& 2.0	&  $-$	& 13	& 4 \\
\hline
HD 28192 b                  	& $-$2 	& 100	& $\sim12$	& 0.118	& 0.10	& 279	&  8.7	&  $-$	& $-$	&  $-$	&  4	& 1 \\
WASP-20 b\tablenotemark{*}  	& $-$24	& 104	& 16	& 0.060	& 0.02	&  90	& 11.2	& 5940	& 1.4	& 1350	&  5	& 1 \\
WASP-174 b\tablenotemark{*} 	& $-$41	& 105	& 16	& 0.055	& 0.00	&  90	& 12.3	& 6400	& 1.3	& 1520	&  3	& 1 \\
HAT-P-67 b\tablenotemark{*} 	& +45  	& 108	& 23	& 0.065	& 0.00	&  90	& 10.5	& 6410	& 2.5	& 1900	& 11	& 3 \\
HD 63454 b                  	& $-$78	& 109	& $\sim13$	& 0.038	& 0.00	&  87	& 10.4	& 4820	& 0.8	&  $-$	& 19	& 5 \\
HD 149026 b\tablenotemark{*}	& +38  	& 109	&  8	& 0.043	& 0.01	& 109	&  8.8	& 6150	& 1.5	& 1630	&  9	& 3 \\
WASP-63 b\tablenotemark{*}  	& $-$38	& 114	& 16	& 0.056	& 0.05	&  90	& 11.7	& 5550	& 1.8	&  $-$	&  4	& 1 \\
HD 33283 b\tablenotemark{a}            & $-$27	& 114	& $\sim13$	& 0.151	& 0.46	& 156	&  8.7	& 6050	& 1.7	&  $-$	&  4	& 1   
\label{subjov}
\enddata
\tablenotetext{*}{Transiting planet}
\tablenotetext{a}{Only two nights needed}
\tablenotetext{}{Super Earths/sub-Neptunes are shown for $M, M \sin i < 17 \, M_\oplus$, super Neptunes/sub-Saturns for $17 \, M_\oplus < M, M \sin i < 95 \, M_\oplus$, and super Saturns/sub-jovians for $95 \, M_\oplus < M, M \sin i < 115 \, M_\oplus$.}
\end{deluxetable*}

\begin{deluxetable*}{lrcccccccccccc}
\tabletypesize{\scriptsize}
\tablecaption{Estimated SNR for $T_\text{eq} \leq 1000 \, K$ Exoplanets with 30-m Class Telescopes}
\tablewidth{0pt}
\tablehead{
\colhead{Planet} & \colhead{Dec} & \colhead{$M, M \sin i$} & \colhead{$R$} & \colhead{$a$}	& \colhead{$e$} & \colhead{$\omega$} & \colhead{$B$mag} & \colhead{$T_*$} & \colhead{$R_*$} & \colhead{$T_\text{eq}$} & \colhead{SNR$_\text{30m}$} \\
& ($^\circ$) & ($M_J$) & ($R_J$) & (au) & & ($^\circ$) & & ($K$) & ($R_\odot$) & ($K$) & (1 week) }
\startdata
HD 20782 b\tablenotemark{a}                      	& $-$29	&  1.3	& $\sim 1.0$	& 1.368	& 0.95	& 143	&  8.0	& 5780	& 1.1	&  250	&  4 \\
HD 80606 b\tablenotemark{*a}     	& +51  	&  4.1	& 1.0	& 0.455	& 0.93	& 233	&  9.8	& 5570	& 1.1	&  400	&  6 \\
GJ 86 b                         	& $-$51	&  4.3	& $\sim 1.0$	& 0.113	& 0.05	& 260	&  6.9	& 5180	& 0.8	&  650	&  5 \\
55 Cnc b                        	& +28  	&  0.8	& $\sim 1.0$	& 0.115	& 0.00	&  86	&  6.8	& 5180	& 0.9	&  720	&  3 \\
HD 162020 b\tablenotemark{a}                     	& $-$40	& 12.1	& $\sim 1.0$	& 0.077	& 0.28	& 333	& 10.1	& 4770	& 0.8	&  720	&  4 \\
HS Psc b                        	& +27  	&  1.5	& $\sim 1.0$	& 0.043	& 0.27	& 176	& 12.7	& 4200	& 0.7	&  780	&  5 \\
HD 130322 b                     	& $-$0 	&  1.1	& $\sim 1.0$	& 0.090	& 0.02	& 181	&  8.8	& 5360	& 0.9	&  800	&  5 \\
TIC 393818343 b\tablenotemark{*a}	& +4   	&  4.3	& 1.1	& 0.129	& 0.61	&   2	& 10.3	& 5760	& 1.1	&  810	&  3 \\
WASP-80 b\tablenotemark{*}      	& $-$2 	&  0.6	& 1.0	& 0.035	& 0.00	&  94	& 13.3	& 4150	& 0.6	&  830	&  3 \\
TIC 46432937 b\tablenotemark{*} 	& $-$15	&  3.2	& 1.2	& 0.021	& 0.00	&  90	& 15.7	& 3570	& 0.5	&  870	&  3 \\
HD 17156 b\tablenotemark{*a}     	& +72  	&  3.2	& 1.1	& 0.163	& 0.67	& 122	&  8.8	& 6080	& 1.5	&  880	&  3 \\
HD 195019 b                     	& +19  	&  3.7	& $\sim 1.0$	& 0.138	& 0.02	& 136	&  7.6	& 5810	& 1.5	&  930	&  3 \\
HD 103720 b                     	& $-$3 	&  0.6	& $\sim 1.0$	& 0.050	& 0.09	& 262	& 10.5	& 5020	& 0.7	&  930	&  7 \\
WASP-69 b\tablenotemark{*}      	& $-$5 	&  0.3	& 1.1	& 0.045	& 0.06	&  90	& 10.9	& 4700	& 0.8	&  960	&  5 \\
HAT-P-20 b\tablenotemark{*}     	& +24  	&  7.3	& 0.9	& 0.036	& 0.02	& 337	& 12.5	& 4590	& 0.7	&  960	&  3 \\
DMPP-2 b                        	& $-$34	&  0.4	& $\sim 1.3$	& 0.066	& 0.08	&  90	&  9.0	& 6500	& 1.8	& 1000	& 12   
\label{eccen}
\enddata
\tablenotetext{*}{Transiting planet}
\tablenotetext{a}{Only two nights needed}
\end{deluxetable*}

\noindent While POLISH2 is able to observe even the planet Venus at the Lick 3-m, such high photon flux requires PMT detector gain to be reduced. This effectively reduces detector QE to prevent damage. Thus, while POLISH2 is able to observe bright exoplanet host stars at essentially any telescope, a bright limit exists for each telescope after which SNR no longer increases with stellar brightness. For 30-m, 8-m, 6.5-m, and 1-m class telescopes, these bright limits are $B$mag $\sim$ 7.8, 5.0, 4.5, and 0.5, respectively. We factor this into our model, which is significant for $\upsilon$ And b at Gemini and ten exoplanets at 30-m class telescopes.

For eccentric planets like GJ 3222b, HD 20782b, and HD 80606b, Equation \ref{snrscaling} must be modified to account for orbital geometry. Specifically, the argument of periastron $\omega$ is key to determining whether the planet passes close enough to the star at the orbital phase conducive to large polarization, and this geometry is included in our model \citep{Kane2010}. We assume all planets in Tables \ref{allplanets} through \ref{eccen} share the same large polarization from small scattering particles as HD 189733b, which is not likely to be a realistic assumption for all exoplanets. However, given that such atmospheric parameters are precisely what a polarimetric campaign is designed to measure, we present the simplest reasonable model to create the target lists. Indeed, any statistically significant non-detection of these exoplanets will provide meaningful constraints on scattering particle properties.

\begin{figure}
\centering
\includegraphics[width=0.47\textwidth]{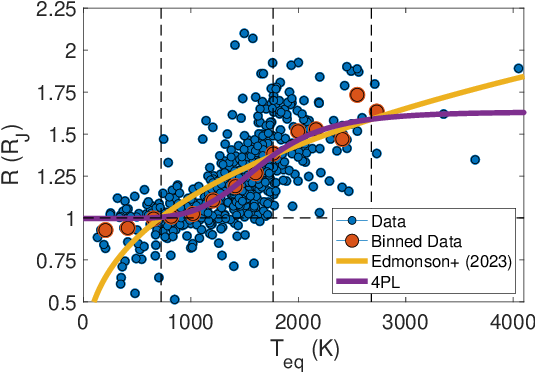}
\caption{Radius vs. equilibrium temperature for transiting jovians in the NASA Exoplanet Archive (blue points), values binned in equilibrium temperature to guide the eye (red points), and $R$-$T$ fits using a power law \citep[Equation \ref{edmond}]{Edmondson2023} and a four-parameter logistic equation from this work (4PL, Equation \ref{logistic}). The power law is not only unphysical for low $T_\text{eq}$, but it overpredicts planetary radius for 725 K $< T_\text{eq} < 1750$ K hot Jupiters (gold curve vs. red, binned data points). The 4PL fit accurately reproduces mean $R$-$T$ measurements (purple curve vs. red points) and naturally accounts for cold jovians such as Jupiter. Dashed vertical lines indicate where the power law and 4PL fits intersect.}
\label{massrad}
\end{figure}

Assuming all planets have the same intrinsically large $B$ band polarization as HD 189733b, we estimate there are 52 other planets detectable with at least $5\sigma$ confidence after one week at Gemini North. Indeed, HD 189733b's SNR only ranks in the 60th percentile of these targets, all of whom have orbital periods shorter than 11 days. The transiting, ultrashort period hot Jupiters KELT-9b, WASP-33b, and HIP 65A b are expected to be easily detectable, or at least are expected to provide stringent constraints on scattering particle radius and composition. Also noteworthy are understudied, non-transiting hot Jupiters such as $\tau$ Boo b, HD 179949b, and $\upsilon$ And b. The general lack of known orbital inclination for non-transiting exoplanets causes two effects, both of which are likely to amplify peak linear polarization, and therefore SNR, over exoplanets in transiting geometries. First, the lack of direct radius measurements requires us to estimate radii using mass-radius-equlibrium temperature ($M$-$R$-$T$) relations. For jovians, Figure \ref{massrad} shows both the following power law relation from the literature \citep[Equation \ref{edmond}]{Edmondson2023, Sousa2024} as well as a four-parameter logistic fit we compute from NASA Exoplanet Archive data (Equation \ref{logistic}):

\begin{subequations}
\begin{align}
\label{edmond} R / R_J & = 1.10 \,T_\text{eq}^{0.35}, \\
\label{logistic} R / R_J & = 1.63 - \frac{0.64}{1+(T_\text{eq}/1650)^{5.29}}.
\end{align}
\end{subequations}

The power law fit overpredicts the radius of transiting hot Jupiters with 725 K $< T_\text{eq} < 1750$ K, which is significant for many of the high SNR jovians in Table \ref{allplanets}. These are of course predictions for individual planets using fits to the transit population, so true radius and SNR may diverge dramatically. For example, the nominal prediction for GJ 3222b's radius ($R = 0.34$ $R_J$) is consistent with the measured radii of some transiting exoplanets sharing GJ 3222b's mass (e.g., Kepler-30b, Kepler-33d, and KOI-142b), while other transiting exoplanets with similar mass to GJ 3222b have significantly smaller radii (e.g., Kepler-10c, Kepler-83c, K2-180b). Given that measured $M \sin i$ is a lower limit to true non-transiting planet mass, estimation of radius using $M$-$R$-$T$ relations will also be a lower limit, at least for sub-jovians due to their non-degenerate equations of state.

\begin{figure*}
\centering
\includegraphics[width=0.99\textwidth]{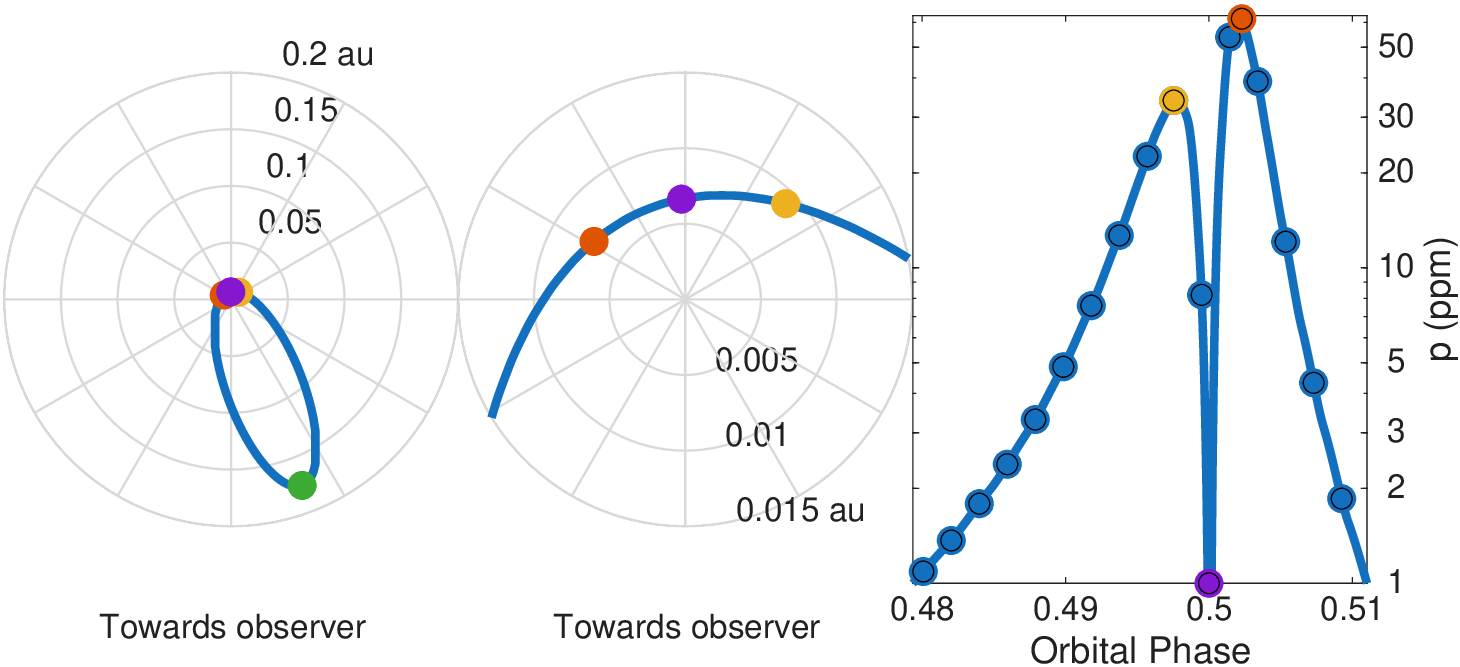}
\caption{GJ 3222b, which is currently the lowest-mass exoplanet amenable to polarimetric scattered light characterization due to its high eccentricity (Tables \ref{allplanets} and \ref{subjov}). \textit{Left}: Top view of the planetary orbit, where the observer is located toward the bottom and the planet orbits counter-clockwise. GJ 3222b's highly eccentric orbit is fortuitously placed such that a highly polarizing geometry occurs for Rayleigh scattering while the planet is close enough to the host star to intercept a detectable fraction of stellar photons. Therefore, GJ 3222b has similar peak linear polarization to an exoplanet on a very close, circular orbit. The purple point indicates periastron and green indicates apastron. \textit{Center}: Zoom of the orbit centered on the star indicating gold and red-orange locations of the linear polarization peaks before and after periastron (purple point), respectively. \textit{Right}: Logarithmic stretch of GJ 3222b's expected Rayleigh-like linear polarization signature, which rapidly peaks (gold point) prior to periastron (purple point), peaks again (red-orange point) after periastron, and rapidly dives back to undetectable levels. Blue points are spaced 30 minutes apart, and the sequence of gold/purple/red-orange points takes 73 minutes.}
\label{gj3222}
\end{figure*}

Low, non-transiting inclinations also provide larger peak linear polarization \citep{Seager2000}, because the polarization orientation of a planet rotates through $\Delta \Theta = 180^\circ$ for face-on orbits but $\Delta \Theta = 0^\circ$ for edge-on orbits. For an exoplanet with no inclination priors, this leads to peak polarization 1.43 times that of a corresponding transiting exoplanet \citep{Lucas2009}, which we account for. Our model suggests that up to eight exoplanets may be detected in scattered light with at least $5\sigma$ confidence after a month of observation each at 1-m class, alt-az telescopes, provided they are sited at appropriate declinations. Indeed, two weeks of $\tau$ Boo b observations at a 1-m alt-az telescope are expected to provide similar SNR to the polarimetric detection of HD 189733b at Gemini North in this work.

Due to fortuitous alignment of its highly eccentric but short period orbit, the non-transiting GJ 3222b = HD 21175b appears to be the lowest-mass exoplanet ($M \sin i = 11.4 \, M_\oplus$) with the highest expected confidence of scattered light detection (Figure \ref{gj3222} and Tables \ref{allplanets} to \ref{subjov}). As a non-transiting planet, however, there currently exists no information about its atmosphere. In contrast, the transiting LTT 9779b ($M = 29 \, M_\oplus$) has been observed by CHEOPS and JWST to harbor an extremely high geometric albedo \citep{Hoyer2023, Coulombe2025}. Indeed, CHEOPS measurements imply a geometric albedo of $A_g = 0.80^{+0.10}_{-0.17}$ \citep[0.35 to 1.10 $\mu$m]{Hoyer2023}, while JWST observes $A_g = 0.79 \pm 0.15$ near the morning terminator and $A_g = 0.41 \pm 0.10$ near the evening terminator \citep[0.6 to 2.8 $\mu$m]{Coulombe2025}. A total of eleven potentially sub-jovian exoplanets ($M, M \sin i < 115 \, M_\oplus$) are expected to be detectable with at least $3\sigma$ confidence in one week of Gemini time, with 73\% of these planets lying in non-transiting geometries (Table \ref{subjov}). Thus, is it possible that non-transiting exoplanets may soon add their weight to our knowledge of sub-jovian atmospheres.

The advent of 30-m class telescopes enables exoplanets down to $M, M \sin i \sim 8 \, M_\oplus$ in transiting (e.g., 55 Cnc e) and non-transiting geometries (e.g., HD 155918b) to be detectable in polarized, scattered light, and it opens the door for scattered light characterization of cooler planets typically too distant from their host star for detection at smaller telescopes. Highly eccentric exoplanets such as HD 20782b and HD 80606b, with $e > 0.9$, $a \ga 0.5$ au, and $T_\text{eq} \la 400\, K$, are detectable during one night per orbit by a brief but large polarization pulse near periastron. Zero-polarization calibration observations may be obtained on one night pre- or post-periastron passage. These exoplanets enable comparison of hot Jupiters with cool atmospheres flash-heated to extreme temperatures. While exoplanet scattered light polarimetry is likely to remain focused on hot Jupiters, there exist certain examples of cooler or smaller exoplanets whose atmospheres may be studied in polarized light.

\section{Conclusion}\label{sec3}

Using the POLISH2 polarimeter at both Gemini North and the Lick 3-m telescopes, we detect linear polarization modulation in $B$ band with peak $\Delta p = 40.9 \pm 7.1$ ppm in the HD 189733 system that is phase-locked to the orbital of the planet. We measure this modulation to be repeatable both in Stokes $q$ and $u$ at two separate telescopes during a seven-year timespan. Furthermore, we measure peak polarization in both Stokes $q$ and $u$ to occur at the expected orbital phase generated by small particles in the atmosphere. Polarimetry favors silicate particles (SiO$_2$ or MgSiO$_3$) with effective radii $r_\text{eff} = 0.038^{+0.047} _{-0.023}$ $\mu$m ($90\%$ confidence) or pure Rayleigh scatterers with geometric albedo $A_g > 0.26$ ($2\sigma$ confidence) over MnS particles with $r_\text{eff} \sim 0.04$ $\mu$m. This is consistent with the JWST detection of $10^{-2.5 \pm 0.7}$ $\mu$m SiO$_2$ particles in HD 189733b's atmosphere \citep{Inglis2024} with a strong disfavor for MnS particles. Combining polarimetry with Hubble secondary eclipse photometry \citep{Evans2013} is difficult, as no model atmosphere consistently explains large polarization and relatively weak secondary eclipse depth.

Direct, scattered light detection of HD 189733b’s high altitude clouds near quarter phase invalidates unocculted starspots as the cause of the planet’s Rayleigh-like transmission spectrum in the blue optical \citep{McCullough2014}, as Rayleigh-like particles are required to provide the large polarization detected. While detected peak linear polarization at the evening terminator is $53.4\% \pm 3.8\%$ larger than the theoretical maximum expected from highly polarizing particles, this planet is known to harbor a super-Rayleigh slope in transmission spectroscopy that implies complexity in the atmosphere. Thus, further modeling is required to investigate the polarimetric effects of such dynamics. We also detect circumstantial evidence that HD 189733b’s peak linear polarization may decrease from the evening terminator to the morning terminator, which may suggest particle condensation on the night side of the planet and particle photoevaporation on the dayside. This hypothesis may be tested with a few additional nights of telescope time. We conclude that the intrinsic, blue optical fractional polarization of HD 189733b is more consistent with Jupiter and Titan than Uranus, Saturn, or Venus, which is the result of small, $\sim 0.04$ $\mu$m particles in the former atmospheres. In contrast, HD 189733b is an order of magnitude more polarized than Venus in the blue optical due to the latter's $\sim 1$ $\mu$m particles.



Not only does exoplanet polarimetry benefit greatly from the large range of scattering angles afforded by short-period orbits, but it also provides significant constraints when multiwavelength observations are utilized. Indeed, while hot Jupiters tend to have low geometric albedos when observed over a broad optical bandpass, the HD 189733b detection in this work shows $B$ band is still sensitive to polarized, scattered light from small particles in hot Jupiter atmospheres. Therefore, a non-detection in $B$ band, as opposed to a non-detection in a broad optical bandpass, would more meaningfully constrain particle properties.

We detect $B$ band circular polarization modulation in the HD 189733 system with peak $\Delta v = 67 \pm 19$ ppm, which is larger than the planetary linear polarization peak of $\Delta p = 40.9 \pm 7.1$ ppm. This modulation appears to be present in our seven-year, two-telescope campaign, and it peaks both at planetary transit and secondary eclipse. Peak circular polarization is many orders of magnitude too large to be intrinsic to the planet, two orders of magnitude too large to be due to instrumental linear-to-circular conversion, and nearly two orders of magnitude too large to be due to the ISM. HD 129333, another BY Dra star like HD 189733, harbors $B$ band circular polarization an order of magnitude larger than observed for HD 189733 \citep{Elias1990}, even accounting for HD 129333's larger starspot filling fraction, which suggests circular polarization to be intrinsic to the HD 189733 host star itself. The clear dependence of HD 189733b orbital phase on measured circular polarization in HD 189733 appears to reinvigorate the debate in the community regarding star-planet interaction in the HD 189733 system \citep{Route2019a}.

The extreme sensitivity and accuracy enabled by large, alt-az telescopes shows that ground-based exoplanet polarimetry may provide crucial constraints on the scatterers in exoplanet atmospheres. To this end, we have commissioned the PHALANX polarimeter (Polarimeter for High Accuracy aLbedos of Asteroids aNd eXoplanets) at the Lick 3-m telescope in 2023 Jan, Jul, and Sep, and PHALANX has been in operation at The Aerospace Corporation Aerotel 1-m alt-az telescope in El Segundo, CA since 2023 Jun. This instrument is a significant upgrade over POLISH2, and it performs part-per-million polarimetry in five optical bandpasses simultaneously.

Since POLISH2 and PHALANX are photon-limited polarimeters, the same accuracy may be obtained on bright stars with PHALANX at the Aerotel 1-m compared to $\sim 4.5$ mag fainter stars at the Gemini North 8-m. This corresponds to the difference in brightness between the ultrashort-period exoplanet hosts KELT-9 and WASP-12. Additionally, if HD 187933b's large $B$ band polarization is mirrored in other known exoplanets, we model up to 52 additional transiting and non-transiting exoplanets to be detectable by their polarized, scattered light with $> 5 \sigma$ confidence at Gemini North or South. The super Earth/sub-Neptune GJ 3222b is the lowest mass exoplanet known ($M \sin i = 11.4 \, M_\oplus$) that is also expected to be detectable in polarimetry due to the geometry of its highly eccentric, $e = 0.93$ orbit. Polarimetric remote sensing is possible for hot sub-jovian or cool jovian exoplanets in the era of 30-m telescopes, though hot Jupiters continue to remain the most attractive targets for the current generation of telescopes. \\

\begin{acknowledgments}

The authors wish to acknowledge scattering calculations performed by Max Michaelis and Sebastian Wolf with the POLARIS code extended to model scattered light from exoplanets.

This work was supported by NASA XRP grant 80NSSC19K1578 and based on POLISH2 observations obtained at both the University of California Observatories/Lick Observatory and the international Gemini Observatory (program ID GN-2018A-C-1), a program of NSF NOIRLab, which is managed by the Association of Universities for Research in Astronomy (AURA) under a cooperative agreement with the U.S. National Science Foundation on behalf of the Gemini Observatory partnership: the National Science Foundation (United States), National Research Council (Canada), Agencia Nacional de Investigaci\'{o}n y Desarrollo (Chile), Ministerio de Ciencia, Tecnolog\'{i}a e Innovaci\'{o}n (Argentina), Minist\'{e}rio da Ciência, Tecnologia, Inovações e Comunicações (Brazil), and Korea Astronomy and Space Science Institute (Republic of Korea). This work was enabled by observations made from the Gemini North telescope, located within the Maunakea Science Reserve and adjacent to the summit of Maunakea.

\end{acknowledgments}

\facilities{Gemini:Gillett (POLISH2), Shane (POLISH2)}

\bibliographystyle{apj}
\bibliography{myrefs}

\end{document}